# Decoupling candidate dual AGN from chance superpositions in the GOTHIC survey via a deep-learning framework

Bhavesh Mukheja[1]★, Snehanshu Saha[2]†, Anwesh Bhattacharya[3], Mousumi Das[4], Françoise Combes[5], Sudhanshu Barway[4]
[1]Department of Physics, Indian Institute of Technology Kharagpur, 721302, West Bengal, India
[2]Department of CSIS, BITS Pilani K K Birla Goa Campus (on lien to Center of AI and SuperComputing, Mahindra University, Hyderabad), Goa, 403726, India
[3]Siebel School of Computing and Data Science, University of Illinois Urbana–Champaign, Champaign, IL 61801, USA
[4]Indian Institute of Astrophysics, Koramangala, Bengaluru 560034, India
[5]Observatoire de Paris, LERMA, College ' de France, PSL University, Sorbonne University, CNRS, Paris, France-75014

Accepted 2026 October 22. Received 2026 October 4; in original form 2026 July 17

## ABSTRACT

Dual active galactic nuclei (DAGN) mark a critical phase in the evolution of merging galaxies and the pairing of supermassive black holes, yet they remain difficult to identify in large imaging surveys because of projection effects and limited spatial resolution. Compact foreground stars and unresolved substructure can mimic dual nuclei through chance superposition, complicating automated detection. We revisit the 46,061 galaxies flagged but rejected as DAGN candidates by the GOTHIC pipeline, primarily because the two nuclei fell within the SDSS fibre aperture or exceeded its separation threshold. We train a supervised deep-learning framework based on the YOLOv11 oriented-bounding-box architecture on annotated SDSS imaging to separate genuine dual nuclei from foreground stellar contaminants and other spurious alignments. The final model attains a validation precision of 0.919, recall of 0.905, and $F_1$ of 0.912, and yields 29,605 dual-nucleus candidates after removing star-dominated and blended detections. Structured visual inspection indicates that 54.5–62% are consistent with genuine dual nuclei, implying $\sim (1.4\text{–}1.8) \times 10^4$ plausible systems. Cross-calibrating the YOLO separation against the deterministic GOTHIC centroid measurement and restricting to the compact regime ($d \leq 6.87''$) gives a conservative subset of $\sim$ 13,672 candidates, reaching calibrated separations of $\sim 0.56''$. Spectroscopy of the most compact ($\leq$ 1 kpc) systems shows spectra dominated by stellar absorption, with no resolved double-peaked emission, so confirmation requires higher-resolution follow-up. The catalogue is a statistically refined list of candidates, not confirmed DAGN. Nonetheless, deep-learning detection substantially reduces contamination and expands the plausible DAGN census.

## 1 INTRODUCTION

Galaxy mergers play a central role in the hierarchical assembly of galaxies and the growth of large-scale structure in the universe (White & Rees 1978; White & Frenk 1991). During these mergers, gravitational torques redistribute angular momentum and drive gas toward galactic centers, often triggering intense star formation and fueling accretion onto supermassive black holes (SMBHs), thereby activating active galactic nuclei (AGN) (Shlosman et al. 1990; Peterson 1997; Hopkins et al. 2010). Depending on their gas content, mergers are commonly classified as *wet* mergers, involving gas-rich galaxies that can produce strong starbursts and ultraluminous infrared galaxies (Sanders & Mirabel 1996; Nandi et al. 2021), or *dry* mergers between gas-poor systems that primarily contribute to the buildup of massive early-type galaxies (Thomas et al. 2005).

An important outcome of galaxy mergers is that as the merger progresses, the nuclear SMBHs of the individual galaxies come closer and finally merge. During this process the central SMBHs are driven toward the nuclear regions through dynamical friction and gravitational torques, eventually forming bound pairs that may coalesce over cosmological timescales (Volonteri et al. 2003; Khan et al. 2016). The in-spiral of SMBHs at parsec scale separations is the primary source of low-frequency gravitational waves in the universe (Salcido et al. 2016). Observationally identifying these systems before coalescence remains a major challenge. Dual active galactic nuclei (DAGN), in which both SMBHs are simultaneously accreting (Das et al. 2018; Rubinur et al. 2018; De Rosa et al. 2019) provide one of the few direct electromagnetic signatures of SMBH pairing and therefore offers critical insight into the final stages of galaxy mergers, black hole growth, and the expected gravitational-wave background (Aggarwal et al. 2019). In recent years triple AGN systems have also been discovered (Pfeifle et al. 2019; Yadav et al. 2021), and suggest the possibility of detecting SMBH clustering, especially within small groups (Foord et al. 2021).

In the literature DAGN are usually defined to be AGN pairs at pro-jected separations below ∼ 10 kpc, although broader definitions may extend to ∼ 50 kpc (Koss et al. 2012; Rubinur et al. 2019a; De Rosa et al. 2019). At even smaller separations (< 100 pc), the SMBHs may form bound systems and are called binary AGNs (BAGNs), representing a later evolutionary stage preceding eventual SMBH coalescence (Kharb et al. 2017a; Ciurlo et al. 2023). DAGN and BAGN systems therefore provide important observational probes of SMBH pairing and the final stages of merger-driven black hole growth. Understanding these systems has become increasingly important in the context of gravitational wave (GW) astrophysics, as the in-spiral of SMBHs is known to be a major source of low-frequency gravitational waves detected by pulsar timing arrays (PTAs) and sources for future space-based observatories such as the Laser Interferometer Space Antenna (LISA) (Abbott et al. 2016; Verbiest et al. 2016; Amaro-Seoane et al. 2017; Hobbs et al. 2010). Cosmological simulations predict that SMBH pairs can persist for several gigayears before coalescence, with some systems stalling at kiloparsec scales or forming long-lived wandering SMBHs in galaxy halos (Tremmel et al. 2017; Ricarte et al. 2021). Consequently, identifying DAGN systems is important for modeling SMBH merger timescales and understanding the demographics of SMBH binaries.

★ E-mail: bhavesh.bm.3000@gmail.com
† E-mail: scibase.snehanshu@gmail.com

Despite their importance, confirmed DAGN systems remain extremely rare, and the number of candidate systems far exceeds the number verified through multi-wavelength or high-resolution follow-up observations (Rubinur et al. 2021). Observational estimates suggest that only a small fraction of AGN reside in dual systems. For example, optical studies of AGN pairs in the Sloan Digital Sky Survey (SDSS) indicate that only ∼ 1.3% of pairs with separations below 30 kpc host dual AGN (Liu et al. 2011), while hard X-ray surveys using Swift-BAT find slightly higher fractions of ∼ 7.5% (Koss et al. 2012). Cosmological simulations similarly predict that active SMBH pairs represent only a small subset of galaxy mergers, with typical fractions below a few percent depending on redshift and separation scale (Volonteri et al. 2016). This apparent scarcity may reflect both intrinsic evolutionary timescales and significant observational challenges.

A variety of observational techniques have been developed to identify DAGN candidates. Early studies relied on double-peaked narrow emission lines as indirect indicators of dual nuclei (Zhou et al. 2004; Rubinur et al. 2019b). However, such spectral features are frequently produced by alternative mechanisms such as rotating gas disks or AGN-driven outflows, leading to high contamination rates (Müller-Sánchez et al. 2015; Kharb et al. 2015, 2021). Multi-wavelength approaches combining X-ray, radio, optical, and infrared observations provide more robust confirmations but are observationally expensive and often limited by angular resolution (Koss et al. 2016; De Rosa et al. 2019). Other techniques include high-resolution near-infrared imaging with adaptive optics (Koss et al. 2018), Gaia-based multi-peak astrometry and variability methods (Mannucci et al. 2022; Hwang et al. 2020), and spectroscopic follow-up of galaxy pairs.

Large imaging surveys, such as SDSS, provide a means to derive large samples of DAGN through identification of galaxy pairs (Ellison et al. 2008). However, to derive nuclei pairs requires more detailed analysis, that include spectroscopic data. A good example is the GOTHIC (Graph-Boosted Iterated Hill Climbing) survey that used a million galaxies and their spectroscopic redshifts to derive a sample of 159 dual/multiple AGN (Bhattacharya et al. 2023). Nevertheless, these methods remain susceptible to selection biases, particularly against compact (< 10 kpc) systems, and are limited by survey constraints such as fiber collisions in large spectroscopic programs like SDSS (Hickox et al. 2009).

Recent advances in machine learning (hereafter ML) have opened new possibilities for addressing these challenges. Deep learning methods, particularly convolutional neural networks (CNNs), have demonstrated remarkable success in galaxy morphology classification and merger detection tasks (Huertas-Company et al. 2015; Domínguez Sánchez et al. 2018; Ribli et al. 2019; Ntampaka et al. 2019). CNN-based models trained on survey data and simulations have achieved accuracies approaching 90% in identifying merger signatures and morphological features (Bottrell et al. 2019; Pearson et al. 2019). These approaches are especially promising for upcoming large-scale surveys such as LSST and Euclid, where automated analysis will be essential for processing vast imaging datasets.

More recently, object-detection frameworks such as the **Y**ou **O**nly **L**ook **O**nce (YOLO) architecture have been adapted for astrophysical applications. These models can efficiently detect and localize sources within images while maintaining high completeness and purity. For instance, the YOLO-CIANNA framework applied to simulated radio survey data achieved detection purities approaching 94% while identifying substantially more sources than traditional methods (Cornu et al. 2024). Similarly, adaptations such as YOLO-CL have demonstrated high completeness and purity in galaxy cluster detection tasks using SDSS data (Grishin et al. 2023; Grishin et al. 2025). These studies highlight the potential of YOLO-based detectors for scalable astronomical source identification and for decoupling DAGNs from chance superpositions.

However, the specific challenge of detecting dual active galactic nuclei in chance superpositions characterized by faint, closely separated, and often overlapping nuclear components remains underexplored using YOLO family of object detection architectures. In particular, distinguishing genuine dual nuclei from compact foreground stars or projection effects presents a significant challenge for automated detection methods.

In this study, we address this problem using a supervised object detection framework based on the YOLO architecture (Redmon et al. 2016), specifically its oriented bounding box implementation (YOLOv11-OBB). Our study focuses on a sample of 46,061 galaxies previously flagged but ultimately rejected as DAGN candidates in the GOTHIC survey primarily because the two nuclei fell within the fibre aperture, and secondarily because their separation exceeded the 40 kpc threshold (Bhattacharya et al. 2023). By training a rotation-aware detector to differentiate between dual nuclei and foreground stars, we aim to refine this ambiguous population and identify previously overlooked DAGN candidates. Throughout, we refer to these systems as dual-nucleus *candidates* rather than confirmed dual AGN. The present method identifies and characterises candidate nuclear pairs from imaging, but spectroscopic confirmation of nuclear activity in both components lies beyond what single-fibre SDSS data can establish (Section 4.7).

Section 2 introduces the YOLO detection framework and describes the dataset used in this study. Section 3 details the staged model development process, including the construction of a baseline detector and the design of a refined two-class rotation-aware model. In Section 4, we present the performance of the final model on the full dataset, identifying 29,605 dual-nucleus candidates after post-processing and overlap filtering. Based on a structured manual inspection strategy, we estimate that ∼ 54.5–62% of these detections correspond to genuine dual-nucleus systems, implying a statistically significant population of $\sim 1.4 \times 10^4$–$1.8 \times 10^4$ plausible candidates. Focusing on the compact-separation regime ($d \leq 6.87''$, the calibrated centroid-to-centroid equivalent of a 10″ YOLO OBB-edge cut), we highlight a conservative subset of ∼ 13,672 physically relevant systems, reaching calibrated centroid-to-centroid separations as small as ∼ 0.56″ (YOLO OBB edge ∼ 4.61″). Section 5 places these results in the context of previous observational and ML studies, and Section 6 summarizes the principal implications of our findings.

[0] It is important to note that in this study we report DAGN candidates, not confirmed DAGNs

## 2 PRELIMINARIES

### 2.1 YOLO for Astronomical Object Detection

You Only Look Once (YOLO) is a family of single-stage deep learning models designed for real-time object detection, first introduced by (Redmon et al. 2016). Unlike two-stage detectors such as R-CNN and its variants which first generate region proposals and subsequently classify them, YOLO formulates object detection as a unified regression problem. Given an input image, the model simultaneously predicts bounding box coordinates, objectness or confidence scores, and class probabilities in a single forward pass through a convolutional neural network (CNN).

CNNs (O'Shea & Nash 2015) have been widely adopted in astrophysics for tasks including galaxy morphology classification, transient detection, and source identification, owing to their ability to learn hierarchical feature representations directly from imaging data (Cornu et al. 2024; Grishin et al. 2023; Grishin et al. 2025). Object detection extends beyond image-level classification by additionally localizing multiple sources within a single frame, a capability that is essential for astronomical images containing crowded fields, overlapping sources, or complex extended structures.

YOLO's single-stage architecture offers several advantages that are particularly relevant for astrophysical applications. First, its global reasoning over the entire image reduces false detections arising from local background fluctuations, a common challenge in low signal-to-noise astronomical data. Second, the model naturally supports multi-scale detection, enabling simultaneous identification of compact sources and extended structures. Finally, its computational efficiency allows for scalable inference on large survey datasets, making it well suited for modern time-domain and wide-field astronomy.

YOLO models operate by partitioning the input image into a grid and predicting bounding boxes relative to each grid cell. Each predicted bounding box is associated with a confidence score, reflecting the model's confidence that the box contains a real source, and a class probability. The quality of a predicted bounding box is typically evaluated using the Intersection over Union (IoU) metric, which quantifies the overlap between predicted and ground-truth regions.

The model outputs for each detection comprise a label file containing the predicted class identifier, confidence score, and oriented bounding box coordinates. To facilitate visual interpretation of the detection results, each class is rendered with a distinct bounding box color in the annotated output images. Specifically, dual nuclei are indicated by dark blue oriented bounding boxes, while foreground star contaminants are indicated by cyan bounding boxes. This color convention is adopted consistently across all figures presenting model predictions in this work.

For clarity, we summarize key terms used throughout this paper:

**Annotation:** Ground-truth bounding box and class label associated with a source.

**Epoch:** One complete pass of the training algorithm over the entire training dataset.

**Confidence score:** The model's estimated probability that a predicted bounding box contains a true source.

**Training dataset**: Annotated images used to optimize model parameters.

**Validation dataset:** A disjoint annotated subset used for hyperparameter tuning and performance monitoring.

**Test dataset:** Held-out data used exclusively for final performance evaluation.

More recently, transformer-based detectors such as DETR and its real-time adaptations (e.g., RT-DETR and related formulations) have gained attention for their end-to-end set-based prediction and reduced reliance on hand-designed components like anchor priors and non-maximum suppression (NMS). In some benchmark comparisons on natural image datasets, optimized transformer detectors have achieved competitive mean average precision with variants of YOLO while eliminating explicit post-processing steps (Parisot & Fernandes 2025). However, these gains often come at the cost of increased computational complexity, slower convergence, and substantially larger model sizes, which can hinder practical deployment, particularly in resource-limited environments or in domains with limited training data. Moreover, transformer-based models can struggle with small and densely packed objects due to global attention patterns that are less sensitive to fine-grained spatial features, whereas YOLO's hierarchical feature maps explicitly encode multi-scale locality.

In this study, we adopt the YOLOv11 architecture (Khanam & Hussain 2024) as the core detection framework. Specifically, YOLOv11n is used as a lightweight baseline model (Section 3.1) , while YOLOv11x with oriented bounding boxes (YOLOv11x-OBB) is employed as the final detection model since DAGNs exhibit no preferred orientation on the sky. Their projected morphologies can appear at arbitrary rotation angles.The choice of YOLOv11 over earlier YOLO variants is motivated by its improved computational efficiency and architectural refinements, which enable faster convergence and enhanced detection performance without compromising accuracy[1].

Our study follows a standard object-detection workflow based on the YOLO framework, adapted to the specific requirements of astronomical imaging data. The overall pipeline consists of (i) annotation of nuclear components in imaging cutouts, (ii) quality control and removal of unsuitable samples, (iii) partitioning of the annotated dataset into training and validation subsets (80:20 split), (iv) model training and convergence monitoring, and (v) application of the trained model to an independent test dataset.

Model performance on the validation dataset was evaluated using standard classification and detection metrics, including Precision ($P$), Recall ($R$), F1-score, Accuracy, and the Positive Likelihood Ratio ($LR^{+}$). Architectural, optimization, and evaluation details of the YOLO framework are described in Appendix A.

### 2.2 Understanding the data

Our sample is derived from our earlier study on galaxy mergers (Bhattacharya et al. 2023), in which an automated image-processing framework **GOTHIC** (Graph-Boosted Iterated Hill Climbing) was developed to identify candidate dual galaxy nuclei in SDSS DR 16. GOTHIC is a deterministic, non-ML algorithm, designed to isolate compact nuclear features through image analysis rather than statistical training.

The GOTHIC pipeline operates through the following stages:

[1] https://docs.ultralytics.com/models/yolo11/#overview

(i) Image normalization and smoothing to suppress large-scale background variations
(ii) Edge detection using a Canny filter applied to 40"×40" image cutouts
(iii) Fitting of a Sérsic light profile to model the host galaxy emission
(iv) Definition of a nuclear search region based on intensity thresholds relative to the Sérsic profile
(v) Identification of local intensity maxima using an iterated hill-climbing algorithm
(vi) Final morphological classification based on peak persistence and spatial separation

A detailed description of the algorithm and its validation is provided in Bhattacharya et al. (2023, Section 5.5).

In the previous study, GOTHIC was applied to a blind sample of 1 million galaxies chosen uniformly at random from SDSS DR16 that have spectroscopic data available. From this parent sample, 104,412 objects were initially flagged as potential merger candidates. Of these, 95,159 sources exhibited consistent nuclear features across multiple photometric bands. Subsequent filtering and conservative morphological vetting reduced this set to 681 high confidence merger systems.

Visual inspection of the 681 selected systems confirms the presence of distinct nuclear components, including a subset of triple-nucleus configurations. Accounting for these cases, the total number of nuclear instances in the confirmed sample is 1,393, which forms the basis of the training and validation data used in this work. Importantly, these multiple nuclei arise naturally from the imaging data and do not rely on artificial augmentation.(Refer: (Bhattacharya et al. 2023))

In addition to the confirmed sample, GOTHIC flagged 46,061 systems that were subsequently rejected from the final catalog, which play a crucial role in this study. These systems were excluded due to a combination of observational and selection-driven constraints inherent to SDSS data and the GOTHIC pipeline.

In the original workflow, systems were required to have spectroscopically distinguishable components (i.e., separate SpecObjIDs without 3″ fiber blending). The surviving systems with projected separations greater than 40 kpc were subsequently excluded to retain physically associated merger candidates. As a result, the rejected sample also contains systems where spectroscopic identifiers may be present, but which were excluded based on separation criteria. We refer the reader to (Bhattacharya et al. 2023) for further details on these selection cuts.

The dominant limitation arises from the 3″ diameter SDSS spectroscopic fiber. When two nuclear components fall within this aperture, only a single spectrum is obtained, preventing the system from being resolved into two spectroscopically independent sources [2]. As a result, such systems lack distinct spectroscopic identifiers (SpecObjID) for both components, making it difficult to confirm their physical association through spectroscopic diagnostics (Guo et al. 2012).

Moreover, additional catalog-level ambiguities contribute to the rejection of these systems. These include photometric blending in crowded fields, misassociation between photometric objects and spectroscopic identifiers, and limitations in resolving compact substructure within a single galaxy using imaging alone. Consequently, the rejected sample is heterogeneous: its members were excluded for several different, and sometimes overlapping, reasons.[3]

**Table 1.** Dataset composition and train–validation split.

| Dataset | Description | Images |
|---|---|---|
| Galaxy mergers | Post-QA confirmed | 1,338 |
| Stars | SDSS sample (post-QA) | 1,741 |
| Bootstrapped mergers | Correct YOLO detections | 40 |
| Total labelled dataset | | 3,119 |
| Training set (80%) | | 2,495 |
| Validation set (20%) | | 624 |
| Testing set | DAGN candidates | 46,021 |

**Table 2.** Class-specific validation performance metrics evaluated at the optimal confidence threshold (C = 0.55).

| Class | P | R | F1 | Acc. | $LR^+$ |
|---|---|---|---|---|---|
| Dual nuclei | 0.967 | 0.803 | 0.877 | 80.3% | 47.2 |
| Foreground star | 0.835 | 0.838 | 0.837 | 83.8% | 16.4 |

Moreover, because GOTHIC operates purely on imaging data, it does not incorporate object-to-objID associations. In regions of high source density or limited spatial resolution, this can lead to ambiguity between genuine dual nuclei and unresolved substructure within a single galaxy. Owing to these uncertainties and the lack of sufficient spectroscopic constraints, these systems were conservatively discarded in the original study.

Crucially, the rejected sample is not astrophysically trivial. It likely contains a mixture of unresolved dual nuclei, single AGNs embedded in complex hosts, foreground stars, and image artifacts. For this reason, we repurpose this set as a challenging test dataset for our ML framework. Its inclusion allows us to evaluate not only detection sensitivity but also the model's ability to suppress false positives and distinguish genuine nuclear multiplicity from confounding morphological features.

## 3 METHODOLOGY

All metadata for the training, validation, and testing samples including `ObjID`, `SpecObjID`, right ascension (RA), and declination (Dec) were obtained from the publicly available GitHub repository[4] associated with Bhattacharya et al. (2023). For each object, RGB cutout images were extracted using the SDSS SkyServer API[5]. Images were downloaded as $120 \times 120$ pixel cutouts with a scale of 0.3″/pixel, a configuration empirically found to provide optimal separation between compact nuclear features and diffuse host-galaxy emission

We adopted a staged model-development strategy to evaluate the suitability of the YOLO object-detection framework for identifying multiple nuclear components in galaxy images. Rather than directly training a complex multi-class model, we began with an intentionally minimal baseline configuration to diagnose intrinsic limitations arising from morphology-driven detection in SDSS survey imaging.

[2] An example: ObjID: 1237650797294125296
[3] Because these catalogue-level ambiguities operate independently of the fibre criterion, a small number of systems in our sample may not satisfy the original > 3″ fibre-separation condition. We retain them deliberately, as resolving such blends is precisely what the present method is designed to test.
[4] https://github.com/anuwu/Blindtest-HQ
[5] https://skyserver.sdss.org/dr16/en/help/docs/api.aspx#imgcutout

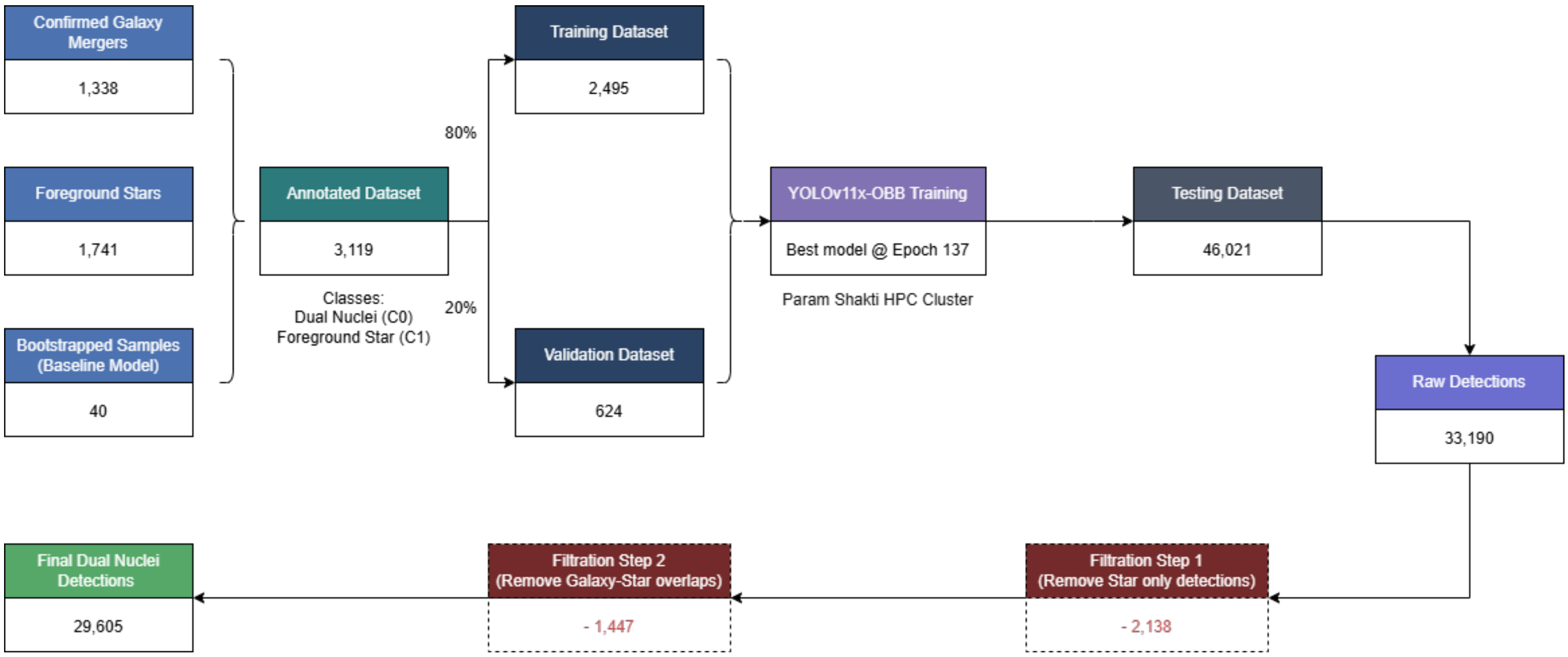

**Figure 1.** Training and inference pipeline of the final YOLO model. Annotated samples are used for supervised training, followed by large-scale inference and post-processing to construct the final dual-nuclei catalog.

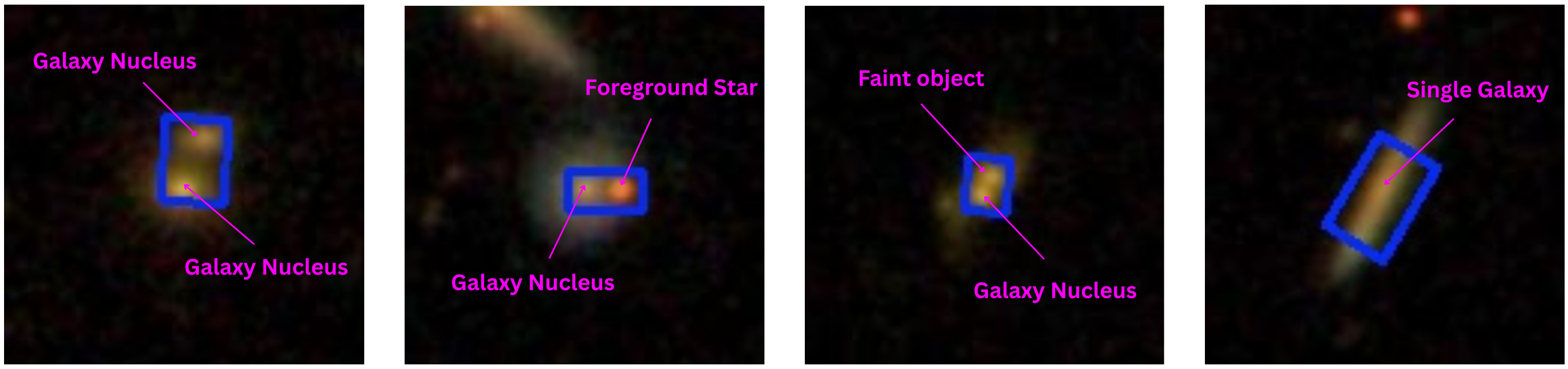

(a) Dual Nuclei (b) Star and galaxy nucleus (c) Faint object & nucleus (d) Single nuclei

**Figure 2.** Examples of the four cases encountered in the 29,605 detections. 2a (`objID: 1237649962452779114`) is Dual Nuclei Detection, 2b (`objID: 1237650371020193968`) is Foreground Star Detection, 2c (`objID: 1237655498126917883`) is Faint Object detection, and 2d (`objID: 1237648722836062301`) is Single Nucleus misclassification. Each panel is a 120 × 120 pixel SDSS cutout at 0.3″ pixel$^{-1}$, spanning 36″ × 36″ on the sky.

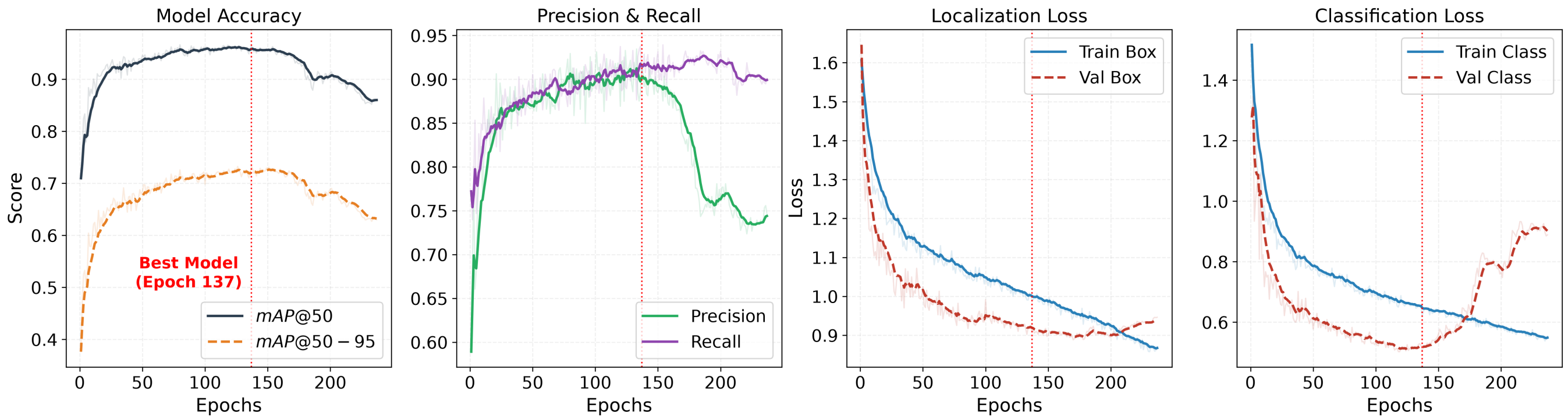

**Figure 3.** Training evolution of the final YOLOv11x–OBB model. Validation mAP, precision, recall, and loss terms are shown as functions of epoch. The vertical dashed line marks the selected best checkpoint (epoch 137).

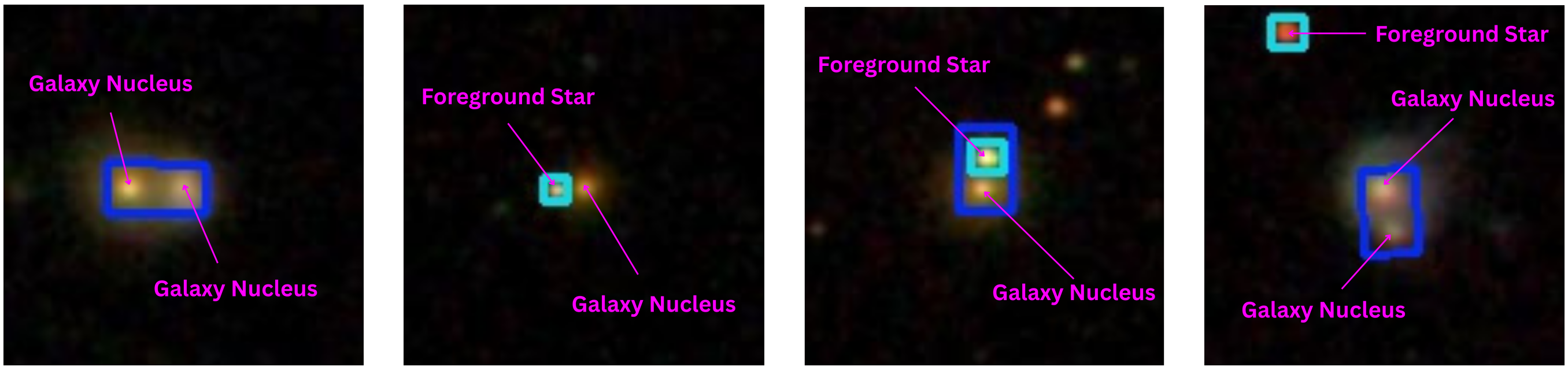

(a) Class 0 (b) Class 1 (c) Overlap (d) Non-overlap

**Figure 4.** Representative detection outcomes. From left to right: dual nuclei (objID: 1237648702973018278), foreground star (objID: 1237648674533934097), overlapping detections (objID: 1237648674534130384), and non-overlapping detections (objID: 1237648703506022589). Each panel is a $120 \times 120$ pixel SDSS cutout at $0.3''$ pixel$^{-1}$, spanning $36'' \times 36''$ on the sky.

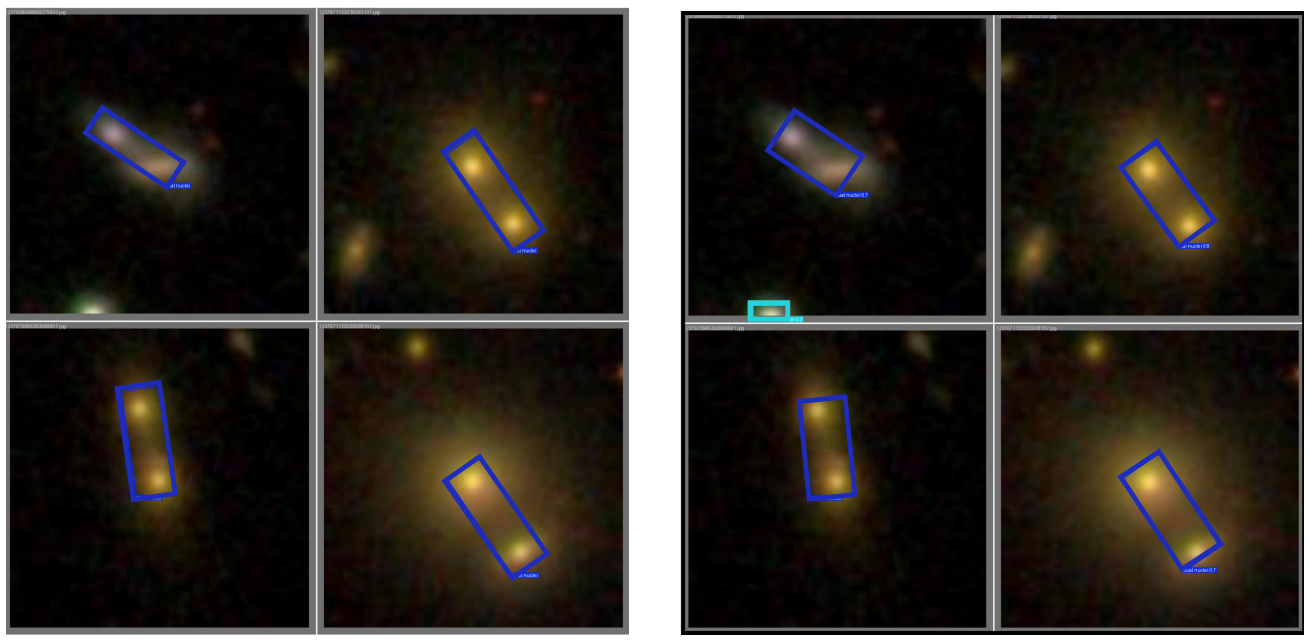

(a) Ground-truth labels (b) Model predictions

**Figure 5.** Example validation images. Left: annotated labels. Right: YOLO detections. Bounding boxes are enlarged for clarity.

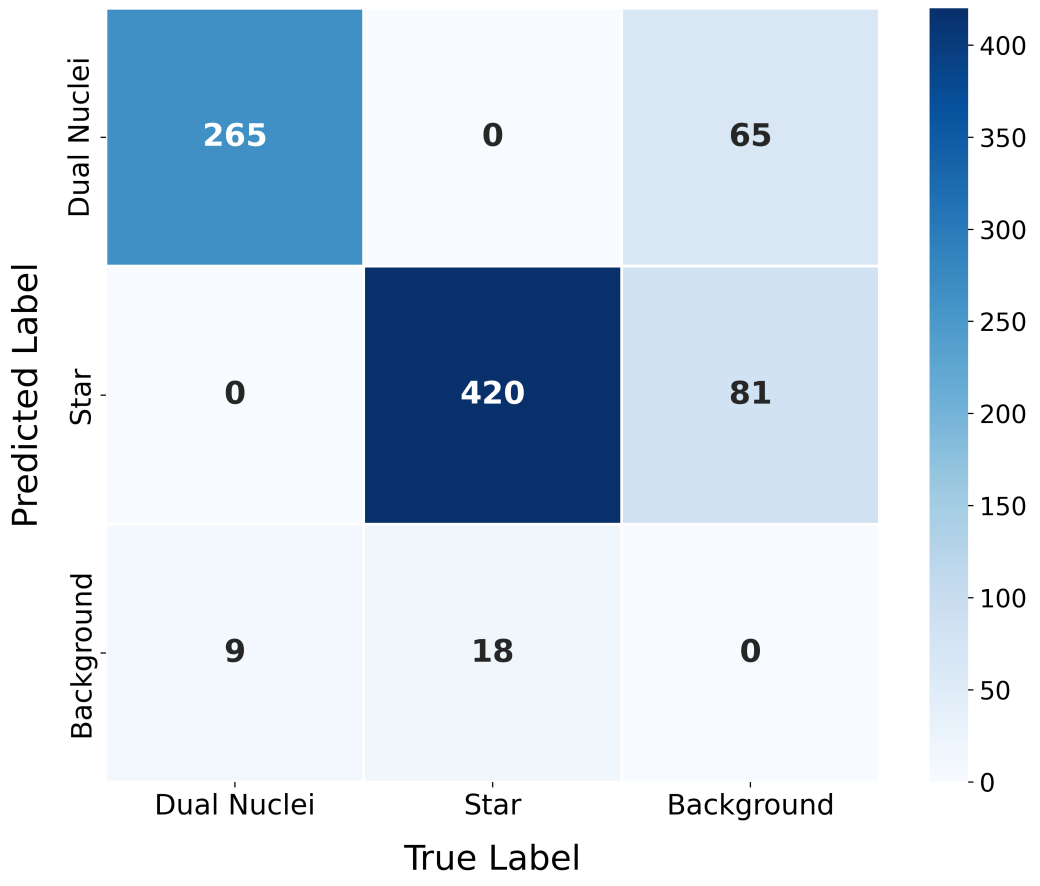

(a) Confusion matrix

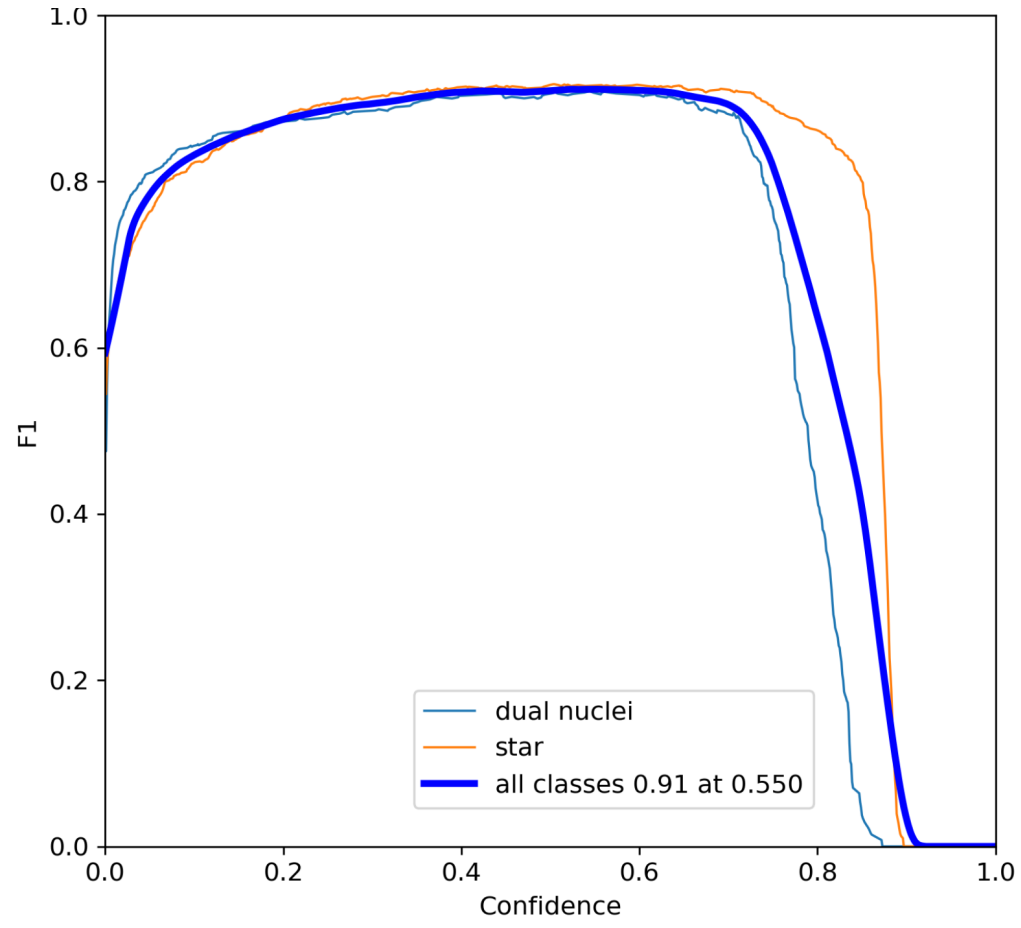

(b) F1 vs. confidence

**Figure 6.** Validation diagnostics of the final YOLOv11x–OBB model. Top: confusion matrix for dual nuclei and stars. Bottom: F1 score vs. confidence, peaking at $F1 = 0.91$ near a threshold of $\sim 0.55$.

**Table 3.** Post-processing and manual inspection results. The manual inspection subsets are mutually exclusive. The post-processing block refers to the full test sample of 46,021 objects; the manual-inspection blocks are drawn from the final catalogue of 29,460 detections with valid separation measurements. The cuts $d \leq 6.87''$ and $d \leq 9.79''$ are the centroid-to-centroid equivalents of YOLO OBB-edge cuts at 10″ and 12.5″ (Equation 2).

| Detection Type | Count | Fraction |
|---|---|---|
| *Full Test Sample* ($N = 46,021$) | | |
| Pure star detections | 2,138 | 4.63% |
| Galaxy–star overlap | 1,447 | 3.13% |
| Final dual-nucleus candidates | 29,605 | 64.07% |
| *Manual Inspection* ($n = 200$, *Full Sample*, $N = 29,460$) | | |
| Dual nuclei | 109 | 54.5% |
| Galaxy nucleus + star | 48 | 24.0% |
| Nucleus + unidentified | 7 | 3.50% |
| Single nucleus (misclassified) | 36 | 18.0% |
| *Manual Inspection* ($n = 100$, $d \leq 6.87''$, $N = 23,173$) | | |
| Dual nuclei | 59 | 59.0% |
| Galaxy nucleus + star | 25 | 25.0% |
| Nucleus + unidentified | 3 | 3.0% |
| Single nucleus (misclassified) | 13 | 13.0% |
| *Manual Inspection* ($n = 100$, $d \leq 9.79''$, $N = 28,494$) | | |
| Dual nuclei | 62 | 62.0% |
| Galaxy nucleus + star | 18 | 18.0% |
| Nucleus + unidentified | 3 | 3.0% |
| Single nucleus (misclassified) | 17 | 17.0% |

### 3.1 Baseline Model

We manually annotated the 1,393 galaxy merger candidates identified by the GOTHIC pipeline using CVAT, an open-source computer vision annotation tool (CVAT.ai Corporation 2023). Following vi-sual inspection, 55 images were excluded from annotation due to extremely poor image quality or unresolved nuclear structure, conditions known to adversely affect object detection performance (Pearson et al. 2019). In this configuration, the model was trained to detect a single object class corresponding to *dual galactic nuclei*. This setup allows us to assess whether a single-class detector can reliably capture nuclear multiplicity based purely on morphological information, while simultaneously revealing dominant sources of false positives inherent to image-based detection. The final annotated dataset therefore consisted of 1,338 images, which were randomly divided into training and validation subsets using an 80:20 split.

**Table 4.** Sensitivity analysis of the inferred dual-nucleus fraction across separation subsets. The separation $d$ is the calibrated centroid-to-centroid angular separation (Equation 2); the cuts $d \leq 6.87''$ and $d \leq 9.79''$ are the centroid-to-centroid equivalents of YOLO OBB-edge cuts at $10''$ and $12.5''$, and select identical systems. Populations $N$ are taken from the final catalogue of 29,460 valid detections. The purity $\hat{p}$ and its 95% binomial confidence interval are derived from the inspected sample of size $n$ and are therefore independent of $N$; projected counts follow from $N\hat{p}$, with the interval obtained by propagating $\hat{p} \pm 1.96\,\sigma_{\hat{p}}$.

| Subset | Population ($N$) | Sample ($n$) | Accuracy ($\hat{p}$) | 95% CI on $\hat{p}$ | Projected Count |
|---|---|---|---|---|---|
| Full sample | 29,460 | 200 | 0.545 | [0.476, 0.614] | $16,056$ $[14,023,\ 18,089]$ |
| $d \leq 6.87''$ | 23,173 | 100 | 0.59 | [0.494, 0.686] | $13,672$ $[11,438,\ 15,906]$ |
| $d \leq 9.79''$ | 28,494 | 100 | 0.62 | [0.525, 0.715] | $17,666$ $[14,956,\ 20,377]$ |

For baseline training, we employed the YOLOv11n architecture initialized with pre-trained YOLOv11n weights. Model optimization and checkpoint selection were performed using the validation set only. No information from the test dataset was used at any stage during training or validation.

### 3.2 Baseline Model Diagnostics and Limitations

The baseline model was trained for a total of 293 epochs, with early stopping triggered automatically by the YOLO training framework rather than by manual intervention (see Appendix A for a detailed description of the training pipeline and stopping criteria). The optimal checkpoint was selected at epoch **212** based on the peak validation-set mAP50-95. The purpose of this training stage was not to establish final model performance, but rather to diagnose systematic failure modes and assess the suitability of a single-class detection framework for the given astrophysical data.

At the optimal checkpoint, the model achieved a validation F1 score of 0.686, with a precision of 0.595 and a recall of 0.809 at a confidence threshold of 0.189. The high recall indicates that the model is sensitive to compact paired structures, while the comparatively lower precision reflects substantial contamination from false-positive detections.

To further assess the discriminative power of positive detections, we evaluated the positive likelihood ratio ($LR^{+}$) which came out to be 0.812. The low $LR^{+}$ value indicates that positive predictions from the baseline model do not provide strong evidence for the presence of genuine dual nuclei. Collectively, these diagnostics demonstrate that while the baseline model can identify candidate nuclear pairs, it lacks the specificity required for reliable application to large, unlabeled survey datasets.

### 3.3 Qualitative Diagnostic Inspection of Baseline Detections

The trained baseline model was applied to the full test set of 46,061 objects, yielding 31,250 candidate detections. To diagnose model behaviour and identify systematic failure modes, we performed a manual inspection of a randomly selected subset of 100 detections. Detections were classified using the SDSS catalogue entry for the secondary component inside the predicted bounding box. If SDSS classifies it as a galaxy, the detection is recorded as dual nuclei (Fig. 2a); if as a star, as a galaxy–star system (Fig. 2b); if a faint source is visible but SDSS has no counterpart for it, as nucleus plus unidentified object (Fig. 2c); and if no distinct secondary component is present, as a single-galaxy misclassification (Fig. 2d). This inspection was conducted solely to inform subsequent model refinement and does not constitute a quantitative evaluation of test-set performance. The inspection revealed three dominant outcomes:

**(i) 47/100** detections correctly localized two distinct galactic nuclei, consistent with SDSS imaging and available metadata (see Fig. 2a).
**(ii) 42/100** detections consisted of one galactic nucleus and one compact point source classified as a star in SDSS. Such cases reflect a well-known degeneracy in SDSS imaging, where bright foreground stars can closely mimic unresolved nuclear components, particularly in the absence of spectroscopic constraints (see Fig. 2b).
**(iii) 5/100** detections corresponded to compact light sources not catalogued by SDSS, potentially indicating faint or unresolved galactic components. These cases are particularly relevant in the context of SDSS fiber-collision limitations and motivate further investigation (see Fig. 2c).

Beyond misclassification, two additional limitations of the baseline configuration were identified. First, the use of axis-aligned bounding boxes failed to adequately capture inclined or rotated nuclear structures, reducing localization fidelity for interacting systems lacking a preferred sky orientation. Second, the lightweight YOLOv11n architecture occasionally produced multiple detections for the same physical system, reflecting limited representational capacity when confronted with complex nuclear morphologies. While the total number of detections remains correct at the image level (i.e., no duplicate images), these effects motivate the adoption of a more expressive architecture and a rotation-aware detection framework.

Collectively, these findings indicate that compact foreground stars constitute the dominant source of false-positive detections in the baseline model, and that rotation invariance and increased model capacity are necessary for robust nuclear identification.

### 3.4 Expanding the Training Set and Redefining the Detection Task

Guided by the diagnostic findings above, we revised both the composition of the training dataset and the formulation of the detection task. In particular, the high incidence of star-induced false positives motivated the explicit inclusion of a *foreground star* class. Incorporating such astrophysical contaminants into the training process has been shown to significantly improve model robustness in related astronomical image-analysis tasks (e.g., Sánchez et al. 2023).

A catalog of stellar sources was constructed via an SQL query exe-cuted on the SDSS SkyServer[6], selecting single isolated 2,000 stars in the same field of view as dual nuclei samples to maintain consistency in observational conditions. Image cutouts were generated using the same pipeline and preprocessing parameters as applied to the merger dataset. Following manual inspection, 259 images were excluded due to faintness, saturation, or ambiguous morphology, yielding a final stellar dataset of 1,741 annotated images.

In addition, we incorporated a limited number of high-confidence merger detections identified during the baseline qualitative inspection. Specifically, 40 of the 47 correctly detected dual-nucleus systems were re-annotated and added to the training set, while the remaining 7 were excluded due to annotation ambiguity. These bootstrapped samples were removed from the test set to avoid evaluation bias. This step modestly expands the training distribution to better reflect the challenging regime encountered in the test sample, particularly in cases where both nuclei fall within the SDSS fiber aperture ($< 3''$). Table 1 summarizes the dataset used in the development of the final model.

### 3.5 Final Model Configuration and Training

The final detection model employs the YOLOv11x architecture with oriented bounding boxes (OBB) and two detection classes: *dual galactic nuclei* and *foreground stars*. The larger YOLOv11x model was selected to provide increased representational capacity and a slower, more stable learning dynamic, which proved essential for resolving closely spaced nuclear components and suppressing spurious detections.

The annotated dataset was split into training and validation subsets using the same 80:20 ratio adopted in the baseline experiment. Training and validation were performed on the Param Shakti high-performance computing cluster at IIT Kharagpur, using identical optimization settings to ensure comparability with the baseline model.

Data augmentation was deliberately kept minimal and was not extended. With only 1,338 annotated merger images, heavier augmentation risked overfitting without adding genuinely new information, and our aim was to keep the training data as close as possible to the imaging conditions of the survey itself. For the same reason we did not supplement the sample with merger systems from other catalogues or surveys, restricting the study to GOTHIC-derived systems keeps the comparison with the original pipeline direct.

The model was trained for 237 epochs (early stopped by YOLO), with the optimal checkpoint identified at epoch 137 based on the peak validation mAP50-95. At this confidence threshold ($C = 0.55$), the model achieved a validation precision of 0.919, recall of 0.905, and an F1 score of 0.912, representing a substantial improvement over the baseline configuration.

### 3.6 Quantitative Evaluation on the Validation Dataset

The robustness and generalization capability of the proposed YOLOv11x-OBB model are assessed through a comprehensive quantitative evaluation on the held-out validation dataset. Figure 6a presents the confusion matrix at the optimal checkpoint (epoch 137), providing a class-wise breakdown of prediction outcomes. The matrix demonstrates strong diagonal dominance for both *dual nuclei* and *star* classes, indicating effective discrimination between compact stellar sources and genuine dual-nucleus systems. Residual off-diagonal entries primarily correspond to visually ambiguous cases, such as compact galaxy–star overlaps, reflecting intrinsic limitations of SDSS imaging rather than systematic model bias.

Figure 3 summarizes the evolution of training and validation losses (box, classification, and distribution focal loss), along with precision, recall, and mean Average Precision (mAP) metrics over the full training schedule. The close tracking between training and validation curves, combined with the absence of late-epoch divergence, indicates stable convergence and limited overfitting. The peak in mAP and F1-score near epoch 137 motivates the selection of this checkpoint as the final model.

To further characterize the operating behavior of the detector, Figure 6b shows the F1-score as a function of confidence threshold for individual classes and their aggregate. The global maximum F1-score of $\sim 0.91$ at a confidence threshold of $\sim 0.55$ reflects a well-balanced trade-off between precision and recall, and supports the threshold adopted for large-scale inference on the full candidate set.

Table 2 reports class-wise performance metrics for the held-out validation set. These should be distinguished from the aggregate model-level metrics quoted in Section 3.5, which summarize overall detector performance at the selected operating threshold. Dual-nuclei detections achieve high precision ($\sim 97\%$), demonstrating effective suppression of star-induced false positives introduced by foreground contamination, while maintaining a recall of approximately 80%. The foreground star class exhibits balanced precision and recall, confirming that the model reliably disentangles compact stellar sources from merger-driven nuclear components.

Taken together, these validation diagnostics provide complementary evidence that the proposed model achieves stable optimization, robust class separation, and reliable generalization within the observational domain, substantially improving upon the baseline architecture.

## 4 RESULTS

Following validation, we applied the final YOLOv11x-OBB model to the full test sample of 46,021 SDSS objects. This inference run yielded a total of 33,190 raw detections, exceeding the baseline model yield. As expected, not all detections correspond to physically meaningful dual-nucleus systems, necessitating post-processing and categorization.

The model assigns detections to two explicit object classes: *dual galactic nuclei* (Class 0) and *foreground stars* (Class 1). Based on the combination and spatial relationship of detected classes within each image, detections naturally fall into four categories:

**(i) Class 0 only**: detections containing exclusively dual-nucleus candidates (Figure 4a).
**(ii) Class 1 only**: detections containing exclusively foreground stars (Figure 4b).
**(iii) Overlapping class detections**: simultaneous Class 0 and Class 1 detections with overlapping bounding boxes, indicating ambiguous or blended sources (Figure 4c).
**(iv) Non-overlapping class detections**: simultaneous Class 0 and Class 1 detections with spatially distinct bounding boxes.(Figure 4d)

### 4.1 Post-processing and False-Positive Mitigation

Our scientific interest lies exclusively in detections containing genuine dual-nucleus systems (Class 0). To suppress residual false positives, we applied a two-stage filtering procedure:

[6] https://skyserver.sdss.org/dr18/SearchTools/sql

**Filtering Step 1**: Images containing only Class 1 (star-only) detections were removed. This step retains images containing pure Class 0 detections, overlapping detections, and non-overlapping mixed detections.

**Filtering Step 2**: Overlapping Class 0 – Class 1 detections were excluded using a geometric intersection test implemented with the `Shapely` library (Gillies et al. 2025). Overlapping bounding boxes are interpreted as ambiguous blends where reliable classification is not possible. The algorithm is further discussed in Appendix C.

Non-overlapping mixed detections, where a dual-nucleus candidate and a star are spatially distinct were retained, as the presence of a foreground star does not compromise the physical interpretation of the nuclear pair.

After filtering, the detection counts are summarized in Table 3.

The final sample of 29,605 dual-nucleus candidates represents a substantial refinement relative to both the baseline YOLO model (31,250 candidates) and the original GOTHIC output, which flagged all 46,021 objects as potential merger candidates. Crucially, the 36% reduction is driven by the explicit identification and removal of foreground-star contaminants, a limitation inherent to purely image-processing based methods such as GOTHIC.

### 4.2 Nature of Residual Detections

Although the final model substantially suppresses star-induced false positives relative to both the baseline configuration and the original GOTHIC pipeline, residual ambiguity remains unavoidable. The retained detections span several physically distinct scenarios:

(i) Bona fide dual-nucleus systems consistent with ongoing or late-stage galaxy mergers.

(ii) A galactic nucleus paired with a foreground star that was not fully disentangled by the classifier.

(iii) A nucleus paired with a faint, low signal-to-noise, or catalog-unmatched source.

(iv) Single galactic nuclei misidentified as dual systems due to internal substructure (e.g., star-forming clumps) or imaging artifacts.

Representative examples of these categories are presented in Figure 2. Notably, similar failure modes were observed in both the baseline YOLO model and the earlier GOTHIC-based analysis; however, their relative frequency is reduced in the final rotation-aware two-class model.

The persistence of these cases reflects intrinsic limitations of SDSS imaging rather than purely algorithmic shortcomings. In particular, surface-brightness sensitivity, limited spatial resolution at higher redshift, projection effects, and catalog incompleteness introduce degeneracies that cannot be fully resolved from single-band morphological information alone. A more detailed breakdown and quantitative assessment of these misclassifications is provided in Section 5.2.

### 4.3 Projected Separation Estimation

To estimate the physical separation of detected nuclear pairs, we adopt a projected-separation approximation. Because spectroscopic redshifts are typically available only for the primary galaxy, we assume that both nuclei lie at the redshift of the primary object. The projected separation is defined as the maximum edge length of the oriented bounding box (OBB) enclosing the detected nuclear pair. This angular separation, $\theta_{\rm arcsec}$, is converted to a physical separation in kpc using

$$s_{\rm kpc} = \frac{\pi \times 10^3\, c}{180 \times 3600 \times H_0}\, \theta_{\rm arcsec}\, z, \tag{1}$$

where $c$ is the speed of light, $H_0$ the Hubble constant, and $z$ the spectroscopic redshift of the primary nucleus. The full procedure, and the values adopted for $c$ and $H_0$, are given in Appendix D.

The OBB edge length is an *upper bound* on the true nuclear separation, since it measures the spatial extent of the enclosing region rather than the centroid-to-centroid distance. Rather than estimating this offset heuristically from the pixel scale, we calibrate it directly. Using the 221 systems in the validation set of the final model for which a separation is available from *both* the YOLO OBB measurement and the deterministic GOTHIC algorithm (Bhattacharya et al. 2023), we treat the two as independent, error-bearing measurements of the same on-sky separation. A method-comparison analysis (Appendix F) yields the calibrated angular relation

$$\theta_{\rm GOTHIC} \;=\; 1.17\,\theta_{\rm YOLO} - 4.83'' \qquad (\text{Pearson } r = 0.91,\ n = 221), \tag{2}$$

and, in physical units, the multiplicative relation

$$s_{\rm GOTHIC} \;=\; 0.66\, s_{\rm YOLO} \qquad (\text{Pearson } r = 0.90,\ n = 221), \tag{3}$$

where the subscripts denote the YOLO OBB-edge and GOTHIC centroid-to-centroid measurements. GOTHIC centroid separations lie on average $3.1''$ below the YOLO OBB edge, with the offset decreasing as separation increases. Applying Equation 2, OBB edge lengths of $\sim$ 8–10$''$ correspond to centroid-to-centroid separations of $\sim$ 5–7$''$.

Throughout the remainder of this work we therefore report GOTHIC-calibrated centroid-to-centroid separations, obtained by applying Equations 2 and 3 to the YOLO measurements, rather than the raw YOLO projected separations. These calibrated values remain *upper bounds* on the true nuclear separation: the longest-edge definition is itself an upper bound, and the YOLO separation enters the calibration as the independent variable. This calibration supersedes the heuristic pixel-scale estimate and places the bulk of the sample in the compact angular regime relevant to late-stage mergers.

A separate caution applies at the smallest separations. Of the net detections with valid calibrated separations, 722 have centroid-to-centroid values of 1.5$''$ or less (Fig. 7a), at or below the typical SDSS seeing limit. These values are obtained by extrapolating Equation 2 below the range of the calibration sample, and are therefore not treated as measurements. We retain these systems in the catalogue and flag them, since if real they would be the most astrophysically interesting members of the sample, but confirming them requires higher-resolution imaging or resolved spectroscopy which is out of scope of this study.

In summary, the pipeline first computes the YOLO projected separation in both arcsec and kpc using the longest-edge algorithm (Appendix D), and then converts these to centroid-to-centroid separations via the YOLO→GOTHIC relations above.

Figure 7 shows the distributions of the calibrated angular and physical separations for the candidate detections produced by the final model. The angular distribution (Fig. 7a) peaks near 5$''$, with a mean of 5.37$''$, a median of 5.19$''$, and a standard deviation of 2.18$''$, extending to a maximum of 22.24$''$. The physical distribution (Fig. 7b) peaks at $\sim$ 7–10 kpc, with a mean of 13.60 kpc, a median of 11.35 kpc, and a standard deviation of 9.13 kpc, extending to a maximum of 91.35 kpc. In both cases an extended tail toward larger separations reflects the presence of widely separated candidates, as well as increasing contamination from non-merger systems.

The two panels are constructed from slightly different subsets, for reasons intrinsic to each quantity. For the angular distribution, the five detections with YOLO edge lengths below 4.13″, the value at which Equation 2 crosses zero, yield non-positive centroid separations and are excluded, leaving 29,600 detections. For the physical distribution, 140 detections are excluded because the SDSS spectroscopic redshift of the primary nucleus is non-positive. A negative value reflects a measurement dominated by peculiar velocity, while a zero entry indicates that SDSS does not provide a reliable redshift. Both cases render Equation 1 ill-defined and are limitations of the SDSS catalogue rather than of the detection method; this leaves 29,465 detections. The two excluded sets are disjoint, so the combined catalogue used for all separation-based analysis in this work comprises 29,605 − 145 = 29,460 detections.

It is important to note that these detections represent raw model outputs and therefore include systems belonging to the categories discussed in Section 4.2, including false positives. Consequently, the histograms should not be interpreted as the intrinsic separation distribution of confirmed dual nuclei, but rather as a convolution of the underlying population and observational selection effects.

Nevertheless, the distributions demonstrate that the model is capable of identifying nuclear components across a broad range of separations, including compact systems approaching the resolution limit of SDSS imaging.

### 4.4 Manual Inspection and Reliability Assessment

Because exhaustive visual inspection of all 29,460 detections is impractical, we performed a structured manual assessment using three independent random samples drawn from different subsets of the detection catalogue. Each system was examined using SDSS imaging and catalogue metadata, treated here as provisional ground truth, following the same classification criteria applied to the baseline inspection (Section 3.3). The inspection results are summarised in Table 3.

The sampling strategy is designed to probe the dependence of detection reliability on angular separation while maintaining statistical independence. All three samples are mutually exclusive. The subsets are defined using cumulative counts derived from the separation distribution shown in Figure 7a:

(i) **Full sample:** 200 randomly selected detections from the complete set of 29,460 candidates.

(ii) **Compact-separation subset** ($d \leq 6.87''$): 100 random detections drawn from 23,173 systems, corresponding to all detections with centroid-to-centroid separations $\leq 6.87''$.

(iii) **Cumulative** $\leq 9.79''$ **subset:** 100 random detections drawn from 28,494 systems, corresponding to cumulative detections with separations $\leq 9.79''$.

These thresholds are the centroid-to-centroid equivalents, obtained via Equation 2, of round-number cuts applied to the YOLO OBB separation: $d \leq 6.87''$ and $d \leq 9.79''$ correspond to YOLO OBB-edge lengths of $10''$ and $12.5''$, respectively. Because Equation 2 is monotonic, the two selections are identical at the object level; we retain the YOLO-defined cuts so that the subsets coincide with those used in the manual-inspection campaign and in the population projections of Table 3. The compact subset therefore isolates the angular regime most relevant to compact dual-nucleus systems, albeit at separations modestly larger than would be inferred from a naive pixel-scale estimate.

The estimated fraction of bona fide dual-nucleus systems is 54.5% for the full sample, 59% for the $d \leq 6.87''$ subset, and 62% for the $d \leq 9.79''$ subset. While these values suggest modest variation across subsets, their statistical consistency is assessed in Section 4.5.

Importantly, these results are obtained without spectroscopic preselection, morphological priors, or synthetic augmentation, and therefore reflect the model's ability to generalise across heterogeneous SDSS imaging conditions.

Within the inspected sample (400 systems), the smallest reliably identified detection has a YOLO OBB edge length of $\sim 4.61''$ (Figure 8b). Applying Equation 2, this corresponds to a calibrated centroid-to-centroid separation of $\sim 0.56''$. It represents an extrapolation and should be regarded as indicative, since the relative scatter of the calibration is largest for the closest pairs. The detection nonetheless demonstrates sensitivity to closely separated nuclear components despite the resolution limits of SDSS imaging.

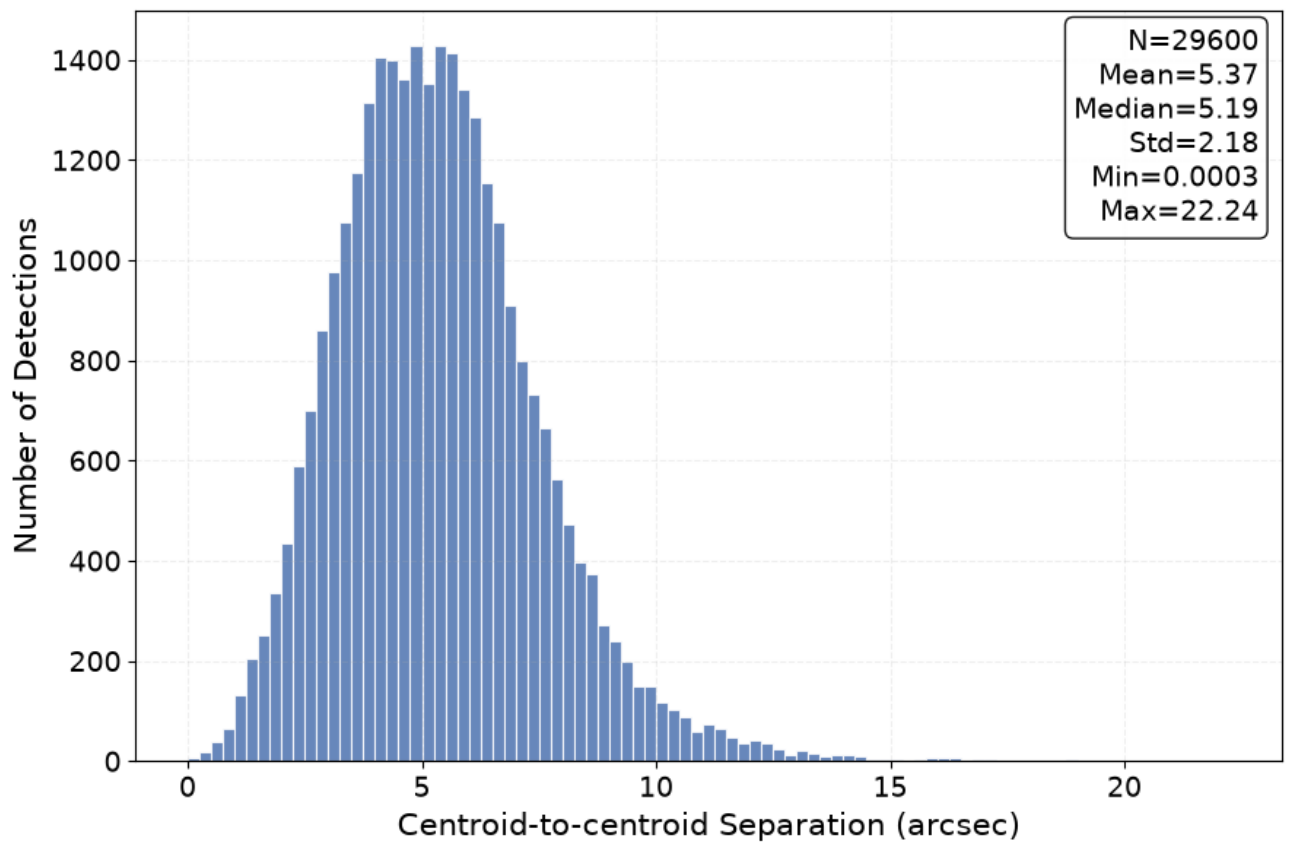

(a) Angular centroid-to-centroid separation.

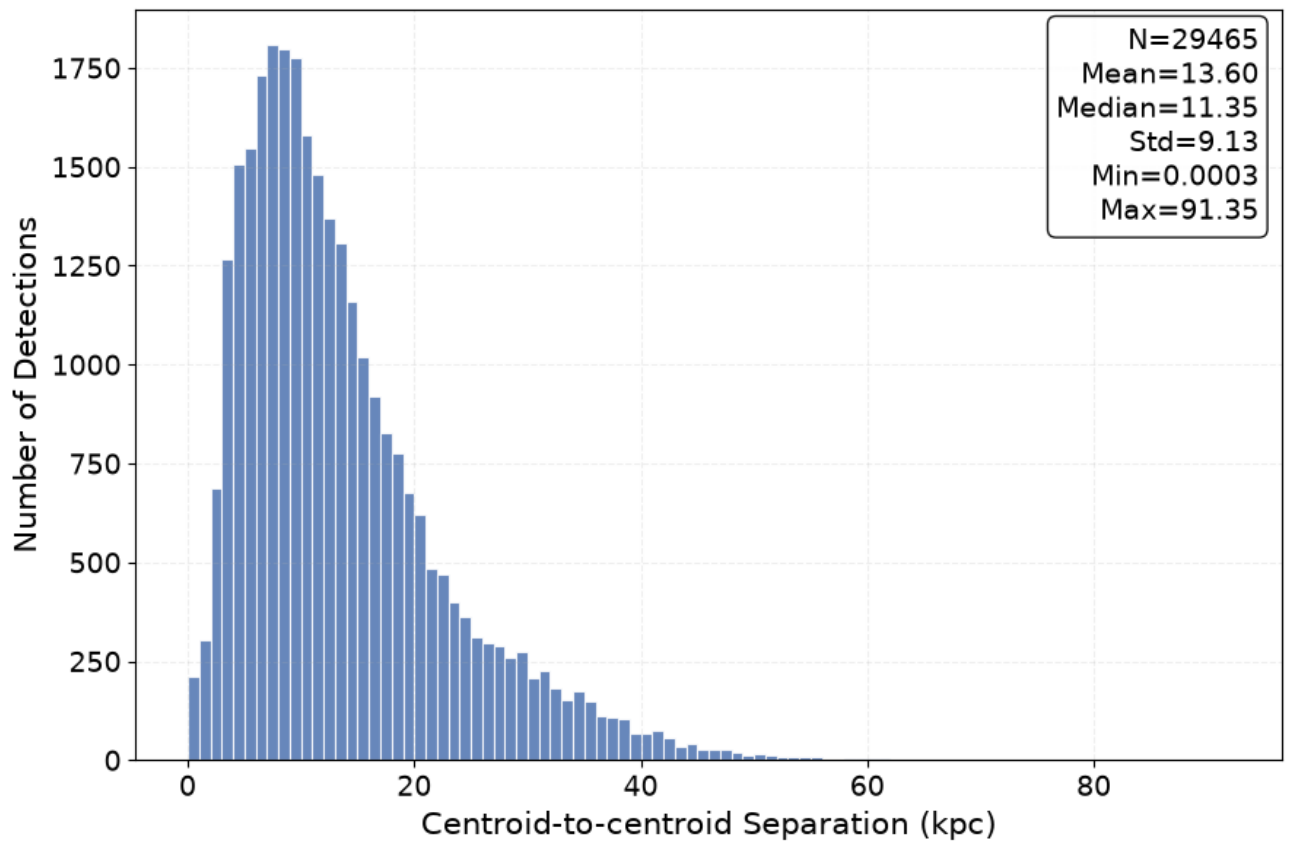

(b) Projected physical separation.

**Figure 7.** Distributions of GOTHIC-calibrated separations for the final-model dual-nucleus detections, obtained by applying the YOLO→GOTHIC relations (Equations 2 and 3). *(a)* Angular separation in arcsec; the five detections with non-positive calibrated separations (YOLO edge $< 4.13''$) are omitted, leaving 29,600. *(b)* Projected physical separation in kpc; the 140 detections with non-positive SDSS redshift are omitted, leaving 29,465. The two excluded sets are disjoint, and the combined catalogue used for all separation-based analysis comprises 29,460 detections. Both distributions show extended tails toward larger separations that include increasing non-merger contamination.

### 4.5 Sensitivity Analysis from Random Sampling

The three independent random samples provide insight into both the stability of the model and the dependence of detection reliability on separation scale. The observed fractions of bona fide dual-nucleus systems (54.5%, 59%, and 62%) are subject to statistical uncertainty arising from finite sample sizes.

To quantify this uncertainty, we model the manual-inspection outcome as a binomial process. Let $X \sim \mathrm{Binomial}(n, p)$ denote the number of true dual-nucleus systems in a sample of size $n$, with estimator $\hat{p} = k/n$. Under the normal approximation, the standard error is given by:

$$\sigma_{\hat{p}} = \sqrt{\frac{\hat{p}(1-\hat{p})}{n}}. \quad (4)$$

A two-sided 95% confidence interval is then:

$$\hat{p} \pm 1.96\,\sigma_{\hat{p}}. \quad (5)$$

Throughout this section $d$ denotes the calibrated centroid-to-centroid separation (Equation 2); the thresholds $d \leq 6.87''$ and $d \leq 9.79''$ are the centroid-to-centroid equivalents of the YOLO OBB-edge cuts at $10''$ and $12.5''$ and, because the calibration is monotonic, select population projections of the same systems. The resulting 95% confidence intervals on the inspected purity are:

- Full sample: $[0.476, 0.614]$
- $d \leq 6.87''$: $[0.494, 0.686]$
- $d \leq 9.79''$: $[0.525, 0.715]$

Because these intervals depend only on the inspected sample size $n$, they are unchanged by the revision of the parent catalogue; only the population-level projections rescale. Applying each purity estimate, with its interval, to the corresponding population in the final catalogue of 29,460 valid detections yields the projected counts summarised in Table 4. Details of the confidence-interval computation are provided in Appendix E.

The confidence intervals exhibit substantial overlap, indicating that the observed differences in estimated purity across subsets are not statistically significant at the 95% confidence level. The variation is therefore consistent with sampling fluctuations rather than reflecting a robust dependence on separation scale.

We therefore do not infer a monotonic relationship between projected separation and detection reliability within the $d \leq 9.79''$ regime. Although systems with larger separations may be expected to exhibit increased contamination from projection effects and unrelated companions, the population $\gtrsim 9.79''$ contains comparatively few detections ($\sim 966$), making a statistically meaningful dedicated sampling analysis impractical within the scope of this study.

The statistical consistency across subsets supports the robustness of the inferred dual-nucleus fraction and justifies its use for population-level extrapolation.

### 4.6 Projected Number of Dual-Nucleus Candidates

The final YOLOv11x-OBB model achieves strong performance on the validation dataset, with a precision of 0.967, recall of 0.803, and an F1 score of 0.877 for the dual-nuclei class. While these metrics quantify model behaviour on labelled data, they do not directly translate into an absolute purity estimate for the unlabelled detection sample.

We therefore rely on the empirical estimates derived from manual inspection (Section 4.4). The three independent samples yield purity estimates of 54.5%, 59%, and 62%, with overlapping confidence intervals (Section 4.5). These values define a statistically consistent range for the fraction of genuine dual-nucleus systems.

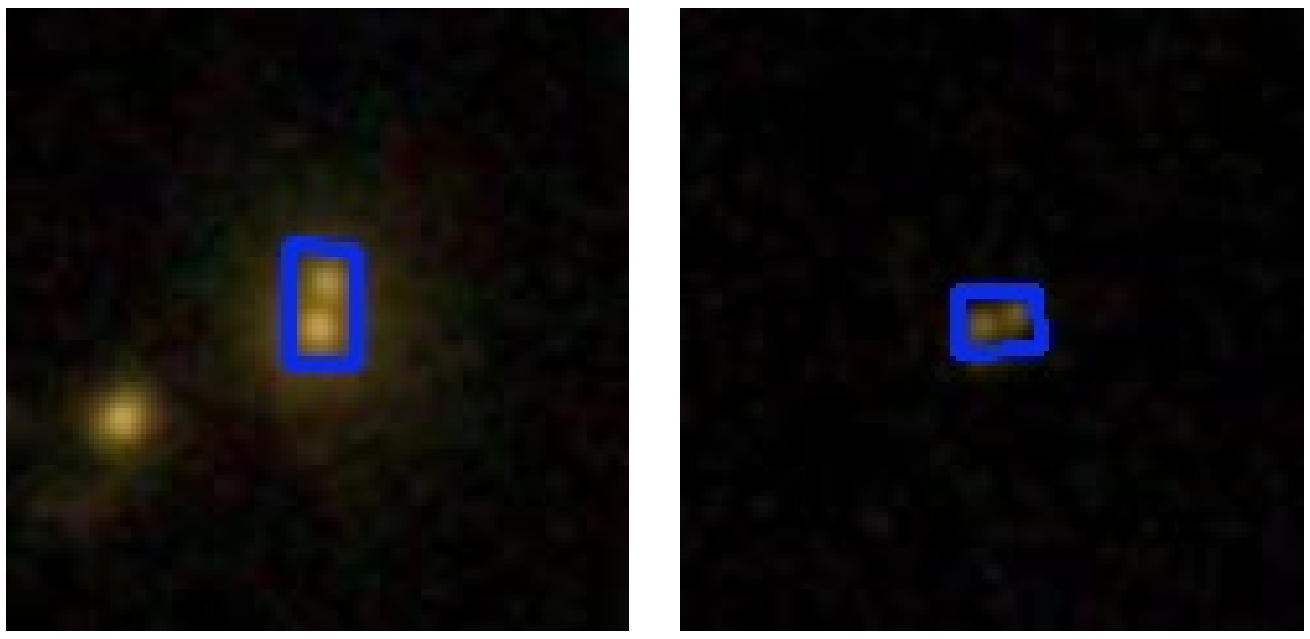

(a) Star–galaxy confusion (b) Least separation estimated

**Figure 8.** Representative edge cases from the inspected sample. Panel a (`objID: 1237651251505528921`) shows a foreground star whose profile mimics a galaxy nucleus. Panel b (`objID: 1237660239778021761`) is the most compact detection in the 400-system inspected sample, with a YOLO OBB-edge separation of $\sim 4.61''$, corresponding to a calibrated centroid-to-centroid separation of $\sim 0.56''$ (Equation 2).

Applying these fractions to the corresponding populations in the final catalogue — 29,460 (full), 23,173 ($d \leq 6.87''$), and 28,494 ($d \leq 9.79''$) — yields projected counts of $\sim$ 16,056, $\sim$ 13,672, and $\sim$ 17,666 systems, respectively. These estimates incorporate sampling uncertainty (Table 4) and should be interpreted as statistical projections rather than definitive classifications.

For practical purposes, the $d \leq 6.87''$ subset provides the most astrophysically meaningful estimate, as it isolates compact centroid-to-centroid separations ($\lesssim 6.87''$, the equivalent of a $10''$ YOLO OBB-edge cut) and is less susceptible to large-scale projection effects. We therefore adopt $\sim$ 13,672 systems as a conservative estimate of likely dual-nucleus candidates within this regime.

For comparison, the GOTHIC pipeline of Bhattacharya et al. (2023) applied to 1 million SDSS galaxies yielded 949 spectroscopic double-nucleus candidates, of which 681 were visually confirmed as genuine nuclei pairs and 159 were ultimately classified as dual AGN through BPT emission-line analysis. These represent successively more stringent confidence levels, and a direct one-to-one comparison with the present catalogue is therefore not straightforward, since our 13,672 systems are statistically inferred dual-nucleus *candidates* rather than individually confirmed pairs or spectroscopically established AGN. Taken at face value, our compact-separation estimate exceeds the GOTHIC confirmed nuclei-pair sample (681) by a factor of $\sim$ 20, and the confirmed dual-AGN sample (159) by close to two orders of magnitude. We stress that these factors compare populations of different confidence: the GOTHIC numbers are the product of individual visual and spectroscopic vetting, whereas the present figure is a population-level projection from sampling-based purity estimates (Section 4.5). The relevant point is not the absolute ratio but that the rejected sample, discarded by GOTHIC precisely because its members fell within the spectroscopic fibre or failed the separation cut, is far from astrophysically empty: it contains a large reservoir of plausible dual-nucleus systems recoverable by morphology-based detection. Follow-up spectroscopic and multi-wavelength observations will be required to establish, for this catalogue, confirmation confidence comparable to the 159 GOTHIC dual AGN.

This expansion reflects the ability of the ML-based framework to suppress dominant image-based contaminants, particularly foreground stars, while retaining sensitivity to compact and closely sepa-rated nuclear systems that are challenging to identify using traditional image-processing methods.

### 4.7 Spectroscopic Analysis of the Most Compact ($\leq$ 1 kpc) Candidates

Manual spectroscopic inspection of all 13,672 candidates in the compact ($d \leq 6.87''$) subset is impractical. We therefore isolate the most physically extreme regime, systems whose calibrated centroid-to-centroid separation satisfies $s_{\rm GOTHIC} \leq 1$ kpc, which contains 212 detections, corresponding to the leftmost bin of the physical-separation distribution (Fig. 7b). The full list of these 212 sub-kiloparsec candidates — SDSS objID, coordinates, and redshift — together with their image cutouts, is provided in the *Supplementary Material*.

This regime is of particular astrophysical interest: at sub-kpc separations the two nuclei are expected to be dynamically bound and approaching the gravitationally hardened binary stage that precedes coalescence and low-frequency gravitational-wave emission (Kharb et al. 2017b). A spectroscopic characterisation of this subset clarifies the physical nature of the detections and the extent to which they can be confirmed with existing survey data.

We emphasise at the outset that a $\leq$ 1 kpc separation is far below the spatial resolution of SDSS imaging, and that the oriented-bounding-box separation is an upper bound (Section 4.3). These systems are therefore not spatially resolved into two nuclei. No separate spectrum exists for the secondary component, so the single-fibre spectrum of the primary is the only direct diagnostic available. A second, more subtle point governs the composition of the subset: because the projected physical separation scales as $s_{\rm kpc} \propto \theta\, z$ (Equation 1), a sub-kpc value can arise either from a genuinely small angular separation or from a low redshift, at which the arcsec-to-kpc scale is small. The $\leq$ 1 kpc subset consequently mixes intrinsically compact angular pairs with low-redshift systems, and is not a pure selection of the closest pairs on the sky.

#### *Redshift reliability*

The conversion to physical units requires a reliable spectroscopic redshift for the primary nucleus. Of the 212 systems, 29 have $z < 10^{-4}$, a value smaller than the general SDSS redshift uncertainty, rendering it impossible to disambiguate genuine compactness from mere redshift measurement error. Thus, we exclude these 29 systems, leaving 183 for spectroscopic analysis.

#### *Method and diagnostic*

Each of the 183 systems was examined using the SDSS interactive spectrum tool. At $\leq$ 1 kpc the two candidate nuclei fall within a single SDSS fibre, yielding one spatially blended spectrum (Guo et al. 2012). In this regime the canonical spectroscopic signature of an unresolved dual active nucleus is a *double-peaked* narrow emission line, in which two narrow-line regions, offset in line-of-sight velocity, produce a split or asymmetric profile in lines such as H$\alpha$ and H$\beta$ (Müller-Sánchez et al. 2015). We accordingly applied a two-stage assessment: 1) whether the spectrum has sufficient signal-to-noise to be analysed, 2) whether the Balmer lines show evidence of a double peak.

Two physical caveats constraint the interpretation of this test, and we apply it conservatively:

**First**, a velocity splitting is resolvable only if the line-of-sight separation exceeds the SDSS instrumental resolution ($R \sim 2000$, corresponding to a velocity FWHM of $\sim 150\,{\rm km\,s^{-1}}$). For a sub-kpc pair the orbital velocities ($\sim 10^2$–$10^3$ km s$^{-1}$ for enclosed masses of $10^{10}$–$10^{11}\,M_\odot$) can in principle exceed this threshold, but random projection of the orbital plane reduces the observable line-of-sight component, so a large fraction of genuine close pairs would present an unresolved / marginally broadened line. A non-detection of a double peak therefore does not a exclude genuine dual nucleus.

**Second**, double-peaked narrow lines are an impure tracer: rotating gas discs, biconical AGN outflows, and jet–ISM interactions can all produce double-peaked profiles in the absence of a second nucleus (Müller-Sánchez et al. 2015; Kharb et al. 2015, 2021). A positive detection is therefore a candidate signature, not a confirmation.

#### *Results*

Of the 183 systems, 40 have spectra too noisy for a meaningful line analysis. The remaining 143 are dominated by spectra in which H$\alpha$ and H$\beta$ appear in *absorption*, together with near-featureless continua. Balmer absorption is the signature of an evolved, predominantly passive stellar population rather than of active accretion. In such spectra the double-peak emission test cannot be applied, since the diagnostic emission lines are absent. In total, 136 of the 143 systems show no indication of double-peaked emission, being either absorption-dominated, featureless, or single-peaked.

Furthermore, 7 systems were initially flagged as possible double-peaked candidates. On detailed inspection however, 6 of these show their Balmer features predominantly in absorption, where a genuine emission-line velocity split cannot be established, and the apparent structure is not a resolved double peak. Only a single system, objID 1237665367972773964 ($z \approx 0.009$; Fig. 9), exhibits clear narrow emission: a strong H$\alpha$ line with the [N II] $\lambda\lambda$6548,6583 doublet, and a narrow H$\beta$ line. This spectrum is nonetheless *single-peaked*: although the H$\alpha$ line is moderately broad, no resolved velocity splitting is present at SDSS resolution, so we cannot attribute it to two distinct narrow-line systems. Its line ratios are physically informative — [N II] $\lambda$6583/H$\alpha \approx 0.4$–$0.5$ together with the strong H$\beta$ favours AGN or composite excitation over pure star formation, although a definitive BPT classification requires the [O III] $\lambda$5007/H$\beta$ ratio, which is not covered here, and the observed H$\alpha$/H$\beta \approx 3.7$ exceeds the Case B value of 2.86, indicating modest internal extinction. Even in this best case, the spectrum provides no evidence for two velocity systems.

#### *Interpretation*

The spectroscopic results yield no confirmed double-peaked emission-line dual-nucleus candidate within the $\leq$ 1 kpc subset. We stress that this negative result needs not necessarily be interpreted as failure, but is rather expected due to the following reasons :
**(i)** Single-fibre blending removes spatial information at these separations.
**(ii)** The SDSS spectral resolution, combined with projection, renders the velocity split of a genuine sub-kpc pair frequently unobservable.
**(iii)** Double-peaked profiles are in any case an ambiguous tracer.

The strong predominance of absorption-line and low-signal-to-noise spectra is consistent with the most compact imaging detections being largely evolved systems, although, since these spectra are not continuum-subtracted, weak nebular emission could be masked by the underlying stellar absorption, consistent with late-stage merger remnants, but also with morphological pairs that do not host active

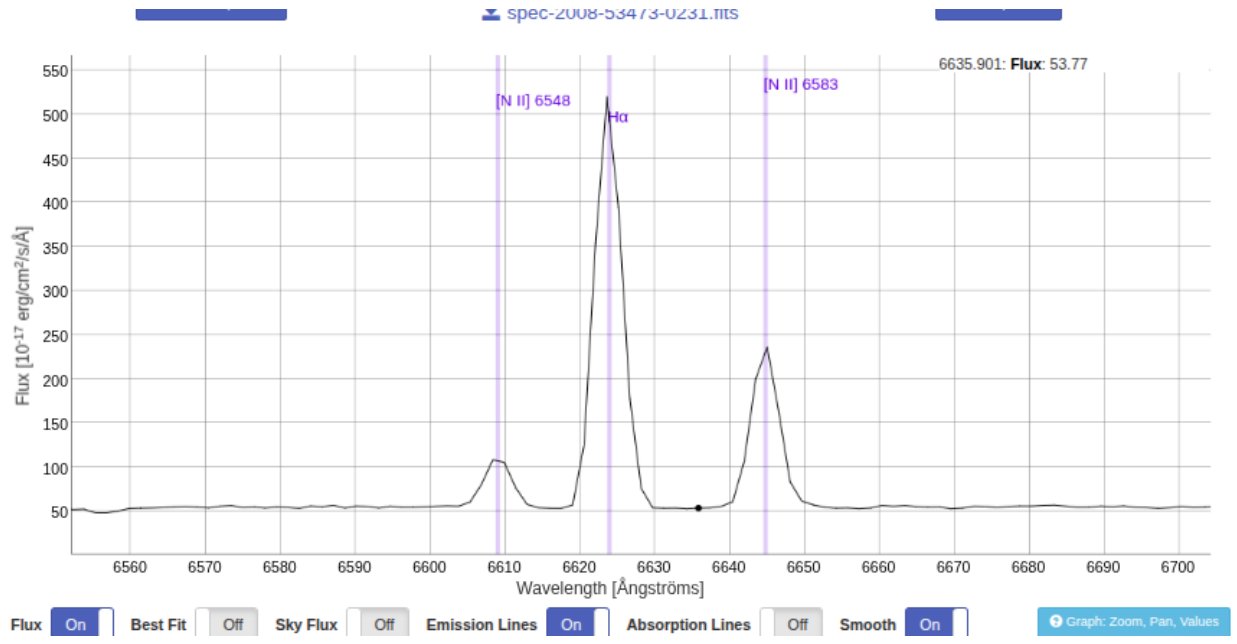

(a) H$\alpha$ + [N II] $\lambda\lambda$6548,6583

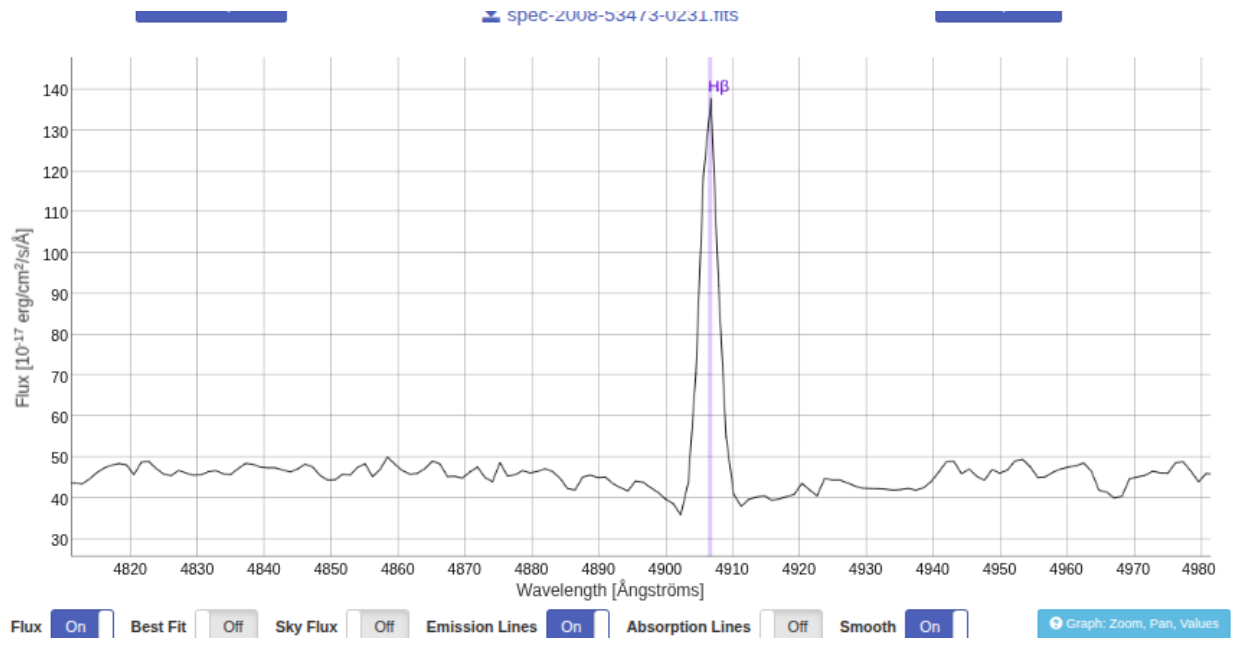

(b) H$\beta$

**Figure 9.** SDSS single-fibre spectrum of objID 1237665367972773964 ($z \approx 0.009$), the only clearly emission-line system in the $\leq$ 1 kpc subset. Panel a shows the narrow H$\alpha$ line and the [N II] $\lambda\lambda$6548,6583 doublet; panel b shows the narrow H$\beta$ line. Both Balmer lines are single-peaked at SDSS resolution, showing no resolved velocity splitting that would indicate two distinct narrow-line systems. Emission lines are marked by the SDSS pipeline.

nuclei. These limitations are intrinsic to SDSS data rather than to the present method, and define the natural boundary of what optical single-fibre spectroscopy can establish for sub-kpc systems. Robust confirmation of dual-nucleus nature in this regime requires spatially resolved spectroscopy (integral-field units such as MaNGA or MUSE), higher spectral resolution to recover small velocity splittings, or independent confirmation at radio (VLBI) and X-ray (e.g. *Chandra*) wavelengths (Kharb et al. 2017b; Koss et al. 2018). The sub-kpc candidates identified here, and in particular the emission-line system objID 1237665367972773964, are well suited to such targeted follow-up.

## 5 DISCUSSION

The primary objective of this study was to reassess the population of candidate dual active galactic nuclei (DAGNs) contained within the sample of 46,061 galaxies previously rejected by the GOTHIC pipeline (Bhattacharya et al. 2023). By introducing a supervised deep-learning framework, we aimed to determine whether a significant fraction of these systems, discarded due to limitations inherent to traditional image-processing techniques, could plausibly host dual nuclear structures consistent with galaxy mergers.

Our methodology followed an explicitly iterative design. We first examined the dataset from the GOTHIC and the statistical composition of the rejected sample, which informed both model selection and label construction. A baseline YOLOv11n detector (Section 3.1) was trained to establish a performance reference and to diagnose dominant sources of false positives and low detection rate, particularly confusion between stellar contaminants and compact galactic nuclei. These insights motivated two key refinements in the final iteration: (i) explicit inclusion of isolated stars as a negative class and (ii) adoption of a rotation-aware detection architecture via YOLOv11x-OBB, which is better suited to resolving closely separated and non-axis-aligned nuclear components.

Unlike several previous studies that rely on aggressive image preprocessing or denoising (e.g., Ackermann et al. 2018; Akhaury et al. 2024), we intentionally operated on minimally processed SDSS images. This choice preserves the observational realism of the survey data and avoids introducing assumptions about the point spread function, which is not uniformly characterized across SDSS fields. As a result, the trained model is exposed to the same noise properties, background variations, and artifacts that would be encountered in large-scale survey applications.

The final model achieves a precision of 0.919 and recall of 0.905 on the validation dataset, indicating effective suppression of dominant false positives while retaining sensitivity to genuine merger candidates. When applied to the full rejected sample, the model identifies 29,605 systems as potential dual-nucleus candidates, of which 29,460 have reliable separation measurements.

To obtain an empirical estimate of the fraction of bona fide dual-nucleus systems, we conducted a structured manual inspection using three mutually exclusive random samples drawn from different separation regimes (Section 4.4). These yield consistent purity estimates of 54.5%, 59%, and 62%, respectively. A statistical analysis using binomial confidence intervals (Section 4.5) shows that these values are mutually consistent within uncertainties, indicating no significant dependence of detection reliability on projected separation within the $d \leq 9.79''$ regime.

Adopting these estimates as a statistically consistent range, we infer that $\sim$ (1.4–1.8) $\times$ $10^4$ systems in the full sample are likely to represent genuine dual-nucleus candidates (Table 4). For astrophysical interpretation, the compact-separation subset ($d \leq 6.87''$, the centroid-to-centroid equivalent of a $10''$ YOLO OBB-edge cut) provides the most physically relevant regime, yielding an estimate of $\sim$ 13,672 candidates.

We emphasize that these values represent statistical projections rather than a definitive census and remain subject to uncertainties arising from projection effects, redshift ambiguity, and morphological degeneracy.

### 5.1 Comparison with the GOTHIC Pipeline

The contrast between the present ML-based framework and the original GOTHIC pipeline highlights the complementary strengths and limitations of the two approaches. GOTHIC is fundamentally an image-processing algorithm that identifies multiple intensity peaks using techniques such as Sérsic modeling, edge detection, and threshold-based segmentation. While highly effective in detecting clear dual nuclei morphologies, it operates independently of object semantics and is therefore vulnerable to confusion between galactic nuclei and unrelated point sources such as foreground stars or dimly lit patches.

In the original study, these limitations necessitated stringent filtering, resulting in the rejection of 46,061 systems, many of which contained compact, ambiguous, or low-contrast structures. The present work builds directly on this output by introducing a learned representation of object morphology. By training on visually confirmed dual-nucleus (merger) systems and explicitly labeled stellar contaminants, the model acquires the ability to distinguish astrophysically meaningful nuclear configurations from visually similar but phys-ically unrelated features. This semantic discrimination constitutes the principal methodological advance over the purely algorithmic peak-finding strategy employed by GOTHIC.

Importantly, the two approaches should not be viewed as competing methodologies. Rather, they represent complementary stages in a scalable discovery pipeline. Image-processing techniques such as GOTHIC remain valuable for rapid, model-independent candidate generation, while ML-based detectors provide a powerful secondary filter capable of reducing false positives and recovering systems missed by heuristic criteria. Applying this combined framework to future surveys with improved spatial resolution and multi-band coverage will be a natural extension of this work.

### 5.2 Analysis of Misclassified Cases in the Final Model

Manual inspection of a randomly selected subset of 200 detections from the 29,460 detections from the final model indicates that approximately 54.5% correspond to plausible dual-nucleus systems. The remaining detections are dominated by misclassified galaxy–star associations (24%), followed by single-galaxy detections (18%), and a small fraction of unidentified compact sources (3.5%). Notably, these categories broadly mirror the dominant false-positive modes reported for the GOTHIC pipeline, although their relative contribution is significantly reduced in the present framework. Notably, the relative proportions of these categories remain broadly consistent across all three sampled subsets (Section 4.4), suggesting that the dominant misclassification modes are not strongly dependent on projected separation within the compact-separation regime explored here.

The galaxy–star misclassifications can be attributed to several factors:

**Limited Training Diversity**: Despite the inclusion of a dedicated stellar class, the number of annotated examples involving close galaxy–star superpositions remains limited. In particular, cases where a bright foreground or background star lies in close angular proximity to a galactic nucleus are underrepresented in the training set, leading to residual ambiguity. Expanding the training dataset to explicitly include such confirmed configurations would likely improve discrimination.

**Star–Galaxy Morphological Degeneracy**: Some stellar sources in SDSS imaging exhibit profiles that deviate from ideal point-spread functions, especially at low signal-to-noise ratios or in crowded fields. These sources can closely resemble compact galactic nuclei, resulting in false dual-nucleus detections (see Fig. 8a).

**Image Quality Constraints**: At higher redshifts, the declining spatial resolution and signal-to-noise ratio of SDSS images exacerbate confusion between galactic cores and nearby point-like sources. This limitation affects both traditional image-processing pipelines and data-driven deep-learning models.

A particularly interesting subset of detections (3.5%) consists of secondary compact sources with no clear counterparts in the SDSS photometric or spectroscopic catalogs. These unidentified objects may arise from several factors, including spectroscopic incompleteness due to fiber collisions, detection limits in crowded or low signal-to-noise regions, or intrinsically faint companions that fall below standard catalog thresholds. Such cases highlight the ability of the model to identify subtle or previously unreported features that are not captured by conventional catalog-based selection. An illustrative example is shown in Fig. 2c, where a secondary component is visible in close proximity to a primary galaxy, exhibiting morphology consistent with a potential merging system. While the physical association of such components cannot be confirmed without photometric and spectroscopic follow-up, these detections represent promising candidates for deeper, higher-resolution or multi-wavelength observations.

Finally, approximately 18% of the detections correspond to single galaxies incorrectly classified as mergers. These false positives are typically associated with irregular or asymmetric morphologies, where star-forming clumps, tidal features, or noise artifacts are misinterpreted as distinct nuclear components. This behavior highlights an inherent limitation of morphology-based detection in low-resolution survey data and points to the need for complementary spectroscopic or kinematic validation.

### 5.3 Comparison with Existing Literature

A wide range of approaches have been proposed for identifying galaxy mergers and dual AGN systems, spanning traditional image-processing techniques, supervised deep learning on observational data, and simulation-driven classification frameworks. Each class of methods operates under different assumptions and constraints, which must be carefully considered when comparing reported performance metrics.

Several observational studies report high classification performance under relatively controlled conditions. For example, Ackermann et al. (2018) achieved precision, recall, and F1-scores exceeding 0.96 using convolutional neural networks trained on visually selected SDSS mergers. Similarly, Pearson et al. (2019) reported an accuracy of 91.5% for a CNN trained directly on SDSS imaging data, significantly outperforming a counterpart trained on EAGLE simulations. However, these datasets are often biased toward morphologically conspicuous mergers with clear tidal features, which can inflate apparent performance and limit generalizability to more ambiguous or compact systems.

Recent simulation-based studies have also demonstrated strong performance. Using Bayesian deep learning applied to IllustrisTNG, Ferreira et al. (2020) reported merger classification accuracies of approximately 90%. The DeepMerge framework Ćiprijanović et al. (2022) achieved substantial performance gains through domain adaptation, improving accuracy from $\sim$ 50% to $\sim$ 79% when transferring from simulations to observational data. Similarly, Bickley et al. (2024) reported an accuracy of $\sim$ 78% for models trained and evaluated on SDSS-like datasets. While these results underscore the power of simulation-informed learning, they often rely on idealized conditions, including noise-free images or telescope-simulated data optimized for specific merger stages.

In contrast, the present study adopts a strictly observational training and validation strategy, using raw SDSS images without preprocessing, denoising, or synthetic augmentation. This choice prioritizes physical realism over optimized metric performance, enabling the model to learn directly from the intrinsic noise, resolution limits, and diversity present in real survey data. As a result, the reported performance metrics are directly representative of deployment in large-scale observational archives, without requiring post hoc domain adaptation.

A broader comparison of merger classification performance across observational studies is summarized in Ackermann et al. (2018, Table 1), while Bickley et al. (2024) provides a complementary overview of simulation-based approaches. Within this landscape, our work is distinguished by its focus on compact dual-nucleus systems drawn from a previously rejected and heterogeneous sample, a regime that remains underexplored in existing literature.

### 5.4 Challenges and Future Work

Despite the encouraging results presented here, several limitations remain that motivate future development. A primary challenge is the lack of spectroscopic redshift measurements for both nuclei in most candidate systems, which prevents definitive confirmation of physical association and the exclusion of chance projections. To estimate physical separations, we adopted a statistically motivated approximation by assuming both components lie at the redshift of the primary galaxy and defining projected separation using the longest edge of the YOLO-OBB bounding box (Appendix D). While appropriate for large-sample statistical analysis, future surveys employing integral-field spectroscopy or targeted follow-up observations will be essential for robust confirmation of true dual AGN systems. Our spectroscopic examination of the most compact ($\leq$ 1 kpc) subset (Section 4.7) makes this concrete. These systems are dominated by absorption-line spectra, and even the single clear emission-line case (objID 1237665367972773964) is single-peaked at SDSS resolution, so neither confirmation nor exclusion of a genuine dual nucleus is possible from single-fibre data alone.

Another limitation arises from the intrinsic quality of SDSS imaging, particularly at higher redshifts and in low signal-to-noise regimes. Under such conditions, galaxies with irregular morphologies, star-forming clumps, or proximity to bright foreground stars can be misclassified as dual nuclei. These effects are not unique to deep learning approaches but reflect fundamental constraints of ground-based wide-field surveys.

Each of the limitations above can be bounded more sharply than we have attempted here with the following analyses. The rate at which the shared-redshift assumption fails can be estimated from field-galaxy surface density, replacing the qualitative caveat above with a number. The expected level of foreground stellar contamination can be predicted from SDSS star counts, which vary strongly with Galactic latitude across our footprint, giving an external check on the star suppression reported in Section 4.1. An emission-line classification of the primary nuclei across the full catalogue, rather than the $\leq$ 1 kpc subset of Section 4.7, would constrain its AGN content.

Several complementary methodologies offer promising directions for future integration. Approaches based on tidal feature detection Walmsley et al. (2021), morphological statistics Wilkinson et al. (2024), simulation-driven classification Margalef-Bentabol et al. (2024), and GAN-based domain transfer Ćiprijanović et al. (2021) could provide additional discriminatory power when combined with nucleus-based detection. Incorporating multi-band photometry or probabilistic redshift estimates may further reduce projection-induced false positives.

Looking ahead, transfer learning presents a natural extension of this work. With appropriate retraining, the proposed framework could be applied to forthcoming surveys such as DESI, LSST, and space-based observatories including JWST. Cross-validation against cosmological simulations (e.g., Illustris, TNG, Horizon-AGN) would further enable systematic studies of selection effects and merger-stage sensitivity.

Finally, the ability to identify compact dual AGN systems at small projected separations is critical for understanding the nature of galaxies in late-stage galaxy mergers (Nehal et al. 2025) and their connection to gravitational wave progenitors De Rosa et al. (2019). As multi-messenger astronomy advances, robust and scalable identification of such systems will become increasingly important. This work demonstrates that deep learning, when trained and interpreted within the observational domain, offers a powerful pathway toward this goal.

## 6 CONCLUSION

In this work, we revisited a previously excluded sample of 46,061 galaxies from Bhattacharya et al. (2023) and demonstrated that supervised deep learning can substantially refine the identification of dual active galactic nucleus (DAGN) candidates within large optical surveys. By developing a YOLO-based object detection framework trained directly on raw SDSS imaging, we introduced an observationally grounded approach that explicitly accounts for foreground stellar contamination and rotational variance through the use of a dedicated star class and oriented bounding boxes.

Starting from a baseline model, qualitative diagnostics informed successive refinements, culminating in a YOLOv11x-OBB configuration that achieved a precision of 0.919, recall of 0.905, and F1-score of 0.912 on the validation dataset, and a precision of 0.967, recall of 0.803, and F1-score of 0.877 for the dual-nuclei class specifically. Unlike purely image-processing methods, the final model learns morphological distinctions between galaxies and stars, enabling it to suppress a dominant source of false positives that affected earlier approaches.

When applied to the full rejected sample, the model produced 29,605 dual-nucleus candidates after post-processing and overlap filtering, resulting in a 36% reduction in candidates. A structured manual inspection strategy based on three mutually exclusive random samples (total $n$ = 400) yields consistent estimates of the fraction of bona fide dual-nucleus systems in the range ∼54.5%–62%. Accounting for statistical uncertainty, these results imply that $\sim 1.4 \times 10^4$–$1.8 \times 10^4$ systems in the sample are likely to represent genuine dual-nucleus configurations. Restricting to the compact-separation regime ($d \leq 6.87''$, the centroid-to-centroid equivalent of a $10''$ YOLO OBB-edge cut), we obtain a conservative estimate of ∼ 13,672 physically relevant candidates. This compact-separation estimate is roughly an order of magnitude (∼ 20×) larger than the 681 visually confirmed nuclei pairs recovered by the original GOTHIC analysis, and far larger still than its 159 spectroscopically confirmed dual AGN. Though we emphasise that our systems are dual-nucleus *candidates* awaiting confirmation, not established dual AGN, so these factors compare populations of differing confidence rather than equivalent detections. Notably, the model is capable of identifying candidates with projected separations as small as $\sim 4.61''$ (YOLO OBB edge; $\sim 0.56''$ in calibrated centroid-to-centroid terms), without relying on spectroscopic preselection, denoising, or handcrafted morphological features.

We emphasize that the resulting catalog should be interpreted as a statistically refined candidate list rather than a definitive census of confirmed DAGNs. Residual contamination from projection effects, irregular morphologies, and unresolved sources remains unavoidable given the limitations of single-band imaging. Nevertheless, by operating entirely within the observational domain and avoiding synthetic augmentation, our framework provides performance metrics that are directly representative of real-world survey deployment. Overall, this study demonstrates that deep learning–based object detection can serve as a powerful and complementary tool to traditional image-processing pipelines for mining large astronomical archives.

## ACKNOWLEDGEMENTS

We are grateful to the referee for a very helpful and constructive report, which improved the clarity of the paper. The first author BM thanks Professor Somnath Bharadwaj (Indian Institute of Technology Kharagpur) for helpful discussions and insightful comments that improved the direction of this research. The authors also acknowledge the Indian Institute of Technology Kharagpur for providing access to the Param Shakti high-performance computing facility, which was used for model training and analysis. MD and SB acknowledge the support of the Department of Science and Technology (DST) grant DST/WIDUSHIA/PM/2023/25(G) for this research.

The authors further thank Professor Meg Urry (Yale University), her research group, and Utsav Akhaury (EPFL) for valuable discussions and insights related to this research.

## DATA AVAILABILITY

The codebase, trained models, preprocessing utilities, post-processing scripts, and instructions required to reproduce the results presented in this work are publicly available through the project repository: https://github.com/BhaveshMukheja/yolo-dagn

A catalogue of the 212 sub-kiloparsec candidates and their SDSS cutouts is available as Supplementary Material of this paper.

The imaging data used in this study originate from the Data Release 16 (SDSS DR16), which is publicly accessible through the SDSS archive. Owing to repository storage constraints and survey data-distribution policies, the full dataset of image cutouts and intermediate products is not hosted directly within the repository. Instead, scripts and instructions are provided to regenerate the dataset and reproduce the analysis pipeline from the original SDSS sources.

Additional derived catalogs generated during this study are available from the corresponding author upon reasonable request.

## REFERENCES

Abbott B. P., et al., 2016, Physical review letters, 116, 061102
Ackermann S., Schawinski K., Zhang C., Weigel A. K., Turp M. D., 2018, Monthly Notices of the Royal Astronomical Society, 479, 415–425
Aggarwal K., et al., 2019, The Astrophysical Journal, 880, 116
Akhaury U., Jablonka P., Starck J.-L., Courbin F., 2024, Astronomy & Astrophysics, 688, A6
Amaro-Seoane P., et al., 2017, arXiv e-prints, p. arXiv:1702.00786
Bhattacharya A., Nehal C., Das M., Paswan A., Saha S., Combes F., 2023, Monthly Notices of the Royal Astronomical Society, 524, 4482
Bickley R. W., Wilkinson S., Ferreira L., Ellison S. L., Bottrell C., Jyoti D., 2024, Mon. Not. R. Astron. Soc., 534, 2533
Bottrell C., et al., 2019, Monthly Notices of the Royal Astronomical Society, 490, 5390–5413
CVAT.ai Corporation 2023, Computer Vision Annotation Tool (CVAT), `https://github.com/cvat-ai/cvat`
Ćiprijanović A., et al., 2022, Mach. Learn. Sci. Technol., 3, 035007
Ciurlo A., et al., 2023, Astronomy & Astrophysics, 671, L4
Cornu D., et al., 2024, Astronomy & Astrophysics, 690, A211
Das M., Rubinur K., Kharb P., Varghese A., Novakkuni N., James A., 2018, Bulletin de la Societe Royale des Sciences de Liege, 87, 299
De Rosa A., et al., 2019, New Astron. Rev., 86, 101525
Domínguez Sánchez H., et al., 2018, Monthly Notices of the Royal Astronomical Society, 484, 93–100
Ellison S. L., Patton D. R., Simard L., McConnachie A. W., 2008, AJ, 135, 1877
Ferreira L., Conselice C. J., Duncan K., Cheng T.-Y., Griffiths A., Whitney A., 2020, Astrophys. J., 895, 115
Foord A., Gültekin K., Runnoe J. C., Koss M. J., 2021, ApJ, 907, 72
Gillies S., van der Wel C., Van den Bossche J., Taves M. W., Arnott J., Ward B. C., others 2025, Shapely, doi:10.5281/zenodo.5597138, `https://github.com/shapely/shapely`
Grishin K., Mei S., Ilić S., 2023, A&A, 677, A101
Grishin K., Mei S., Ilic S., Aguena M., Boutigny D., Paturel M., 2025, Astronomy & Astrophysics, 695, A246
Guo H., Zehavi I., Zheng Z., 2012, The Astrophysical Journal, 756, 127
Hickox R. C., et al., 2009, ApJ, 696, 891
Hobbs G., et al., 2010, Classical and Quantum Gravity, 27, 084013
Hopkins P. F., Younger J. D., Hayward C. C., Narayanan D., Hernquist L., 2010, Monthly Notices of the Royal Astronomical Society, 402, 1693
Huertas-Company M., et al., 2015, The Astrophysical Journal Supplement Series, 221, 8
Hwang H.-C., Shen Y., Zakamska N., Liu X., 2020, Astrophys. J., 888, 73
Khan F. M., Fiacconi D., Mayer L., Berczik P., Just A., 2016, ApJ, 828, 73
Khanam R., Hussain M., 2024, YOLOv11: An Overview of the Key Architectural Enhancements (`arXiv:2410.17725`), `https://arxiv.org/abs/2410.17725`
Kharb P., Das M., Paragi Z., Subramanian S., Chitta L. P., 2015, ApJ, 799, 161
Kharb P., Lal D. V., Merritt D., 2017a, Nature Astronomy, 1, 727
Kharb P., Subramanian S., Vaddi S., Das M., Paragi Z., 2017b, The Astrophysical Journal, 846, 12
Kharb P., Subramanian S., Das M., Vaddi S., Paragi Z., 2021, ApJ, 919, 108
Koss M., Mushotzky R., Treister E., Veilleux S., Vasudevan R., Trippe M., 2012, The Astrophysical Journal Letters, 746, L22
Koss M. J., et al., 2016, The Astrophysical Journal, 825, 85
Koss M. J., et al., 2018, Nature, 563, 214
Liu X., Shen Y., Strauss M. A., Hao L., 2011, The Astrophysical Journal, 737, 101
Mannucci F., et al., 2022, Nat. Astron., 6, 1185
Margalef-Bentabol B., et al., 2024, Astronomy & Astrophysics, 687, A24
Müller-Sánchez F., Comerford J. M., Nevin R., Barrows R. S., Cooper M. C., Greene J. E., 2015, The Astrophysical Journal, 813, 103
Nandi S., Das M., Dwarakanath K. S., 2021, MNRAS, 503, 5746
Nehal C. P., Das M., Barway S., Combes F., Biswas P., Bhattacharya A., Saha S., 2025, MNRAS, 544, 4208
Ntampaka M., et al., 2019, Astrophys. J., 876, 82
O'Shea K., Nash R., 2015, An Introduction to Convolutional Neural Networks (`arXiv:1511.08458`), `https://arxiv.org/abs/1511.08458`
Parisot O., Fernandes D. R., 2025, arXiv preprint arXiv:2508.09831
Pearson W. J., Wang L., Trayford J. W., Petrillo C. E., van der Tak F. F. S., 2019, Astronomy & Astrophysics, 626, A49
Peterson B. M., 1997, An Introduction to Active Galactic Nuclei
Pfeifle R. W., et al., 2019, ApJ, 883, 167
Redmon J., Divvala S., Girshick R., Farhadi A., 2016, You Only Look Once: Unified, Real-Time Object Detection (`arXiv:1506.02640`), `https://arxiv.org/abs/1506.02640`
Ribli D., Pataki B. Á., Zorrilla Matilla J. M., Hsu D., Haiman Z., Csabai I., 2019, MNRAS, 490, 1843
Ricarte A., Tremmel M., Natarajan P., Zimmer C., Quinn T., 2021, Monthly Notices of the Royal Astronomical Society, 503, 6098–6111
Rubinur K., Das M., Kharb P., 2018, Journal of Astrophysics and Astronomy, 39, 8
Rubinur K., Das M., Kharb P., 2019a, MNRAS, 484, 4933
Rubinur K., Das M., Kharb P., 2019b, Monthly Notices of the Royal Astronomical Society, 484, 4933
Rubinur K., Kharb P., Das M., Rahna P. T., Honey M., Paswan A., Vaddi S., Murthy J., 2021, MNRAS, 500, 3908
Salcido J., Bower R. G., Theuns T., McAlpine S., Schaller M., Crain R. A., Schaye J., Regan J., 2016, MNRAS, 463, 870
Sánchez H. D., et al., 2023, Mon. Not. R. Astron. Soc.
Sanders D. B., Mirabel I. F., 1996, Annu. Rev. Astron. Astrophys., 34, 749
Shlosman I., Begelman M. C., Frank J., 1990, Nature, 345, 679
Thomas D., Maraston C., Bender R., De Oliveira C. M., 2005, The Astrophysical Journal, 621, 673
Tremmel M., Karcher M., Governato F., Volonteri M., Quinn T. R., Pontzen

A., Anderson L., Bellovary J., 2017, Mon. Not. R. Astron. Soc., 470, 1121
Verbiest J. P. W., et al., 2016, MNRAS, 458, 1267
Volonteri M., Haardt F., Madau P., 2003, ApJ, 582, 559
Volonteri M., Dubois Y., Pichon C., Devriendt J., 2016, Monthly Notices of the Royal Astronomical Society, 460, 2979
Walmsley M., et al., 2021, Monthly Notices of the Royal Astronomical Society, 509, 3966–3988
White S. D., Frenk C. S., 1991, Astrophysical Journal, Part 1 (ISSN 0004-637X), vol. 379, Sept. 20, 1991, p. 52-79. Research supported by NASA, NSF, and SERC., 379, 52
White S. D., Rees M. J., 1978, Monthly Notices of the Royal Astronomical Society, 183, 341
Wilkinson S., Ellison S. L., Bottrell C., Bickley R. W., Byrne-Mamahit S., Ferreira L., Patton D. R., 2024, Monthly Notices of the Royal Astronomical Society, 528, 5558
Yadav J., Das M., Barway S., Combes F., 2021, A&A, 651, L9
Zhou H., Wang T., Zhang X., Dong X., Li C., 2004, The Astrophysical Journal, 604, L33
Ćiprijanović A., et al., 2021, Monthly Notices of the Royal Astronomical Society, 506, 677–691

## APPENDIX A: YOLO ARCHITECTURE

YOLO (You Only Look Once) formulates object detection as a single-stage, dense prediction problem in which localization and classification are jointly optimized within a single forward pass Redmon et al. (2016).

### A1 Network Formulation

Let an input image be denoted by

$$\mathbf{I} \in \mathbb{R}^{H \times W \times 3}. \tag{A1}$$

The YOLO network represents a parametric mapping

$$f_\theta : \mathbf{I} \to \mathbf{Y}, \tag{A2}$$

where $\theta$ denotes the learnable parameters and $\mathbf{Y}$ is the multi-scale detection tensor containing bounding box parameters, objectness scores, and class probabilities.

Modern YOLO architectures consist of three principal components:

(i) **Backbone:** A fully convolutional feature extractor (CSP/C3-style blocks in YOLOv11) that generates hierarchical feature maps at multiple spatial resolutions.
(ii) **Neck:** A feature pyramid and path-aggregation network (FPN/PAN) that fuses low-level spatial information with high-level semantic features.
(iii) **Detection Head:** A set of prediction layers operating at multiple scales, each regressing bounding box geometry, objectness, and class probabilities.

For each spatial location in a feature map, the detector predicts:

$$\hat{\mathbf{y}} = (\hat{\mathbf{b}}, \hat{o}, \hat{\mathbf{p}}) \tag{A3}$$

where:

- $\hat{\mathbf{b}}$ represents bounding box parameters,
- $\hat{o} \in [0, 1]$ is the objectness score,
- $\hat{\mathbf{p}} \in [0, 1]^C$ is the vector of class probabilities for $C$ classes.

### A2 Oriented Bounding Box (OBB) Parameterization

In standard YOLO models, bounding boxes are axis-aligned and parameterized as $(x, y, w, h)$. In contrast, YOLOv11–OBB extends this formulation to explicitly model object rotation.

Each object is represented by four ordered corner points:

$$\mathbf{b} = (x_1, y_1, x_2, y_2, x_3, y_3, x_4, y_4), \tag{A4}$$

with $(x_i, y_i) \in [0, 1]$ denoting normalized image coordinates. This representation allows the model to learn arbitrary quadrilateral geometries, capturing inclined, elongated, or asymmetric nuclear structures.

Formally, the regression task becomes:

$$\hat{\mathbf{b}} = f_\theta(\mathbf{I}), \tag{A5}$$

where $\hat{\mathbf{b}} \in \mathbb{R}^8$ predicts the rotated bounding box coordinates.

This parameterization is particularly advantageous for astronomical imaging, where galaxy mergers exhibit no preferred orientation on the sky and are subject to projection effects.

### A3 Loss Function

The training objective minimizes a composite detection loss:

$$\mathcal{L}_{total} = \lambda_{box}\mathcal{L}_{box} + \lambda_{cls}\mathcal{L}_{cls} + \lambda_{obj}\mathcal{L}_{obj}, \tag{A6}$$

where:

- $\mathcal{L}_{box}$ is an IoU-based regression loss (e.g., GIoU/DIoU) between predicted and ground-truth OBBs,
- $\mathcal{L}_{cls}$ is a binary cross-entropy loss over class probabilities,
- $\mathcal{L}_{obj}$ is the objectness confidence loss.

For an IoU-based regression term:

$$\mathcal{L}_{box} = 1 - \mathrm{IoU}(\hat{\mathbf{b}}, \mathbf{b}), \tag{A7}$$

where IoU is computed over oriented polygonal regions.

The network parameters are optimized via stochastic gradient descent:

$$\theta_{t+1} = \theta_t - \eta \nabla_\theta \mathcal{L}_{total}, \tag{A8}$$

with learning rate $\eta$.

### A4 Model Selection and Transfer Learning

We employ **YOLOv11x–OBB**, the highest-capacity model in the YOLOv11 series. The model was initialized with pretrained weights from the DOTA-v1.5 rotated object detection dataset, providing strong prior knowledge for oriented localization tasks.

This transfer learning approach improves convergence stability and generalization, particularly in limited-data regimes common in astronomical applications.

### A5 Training and Validation Cycle

Supervised training of YOLO requires annotated datasets consisting of images paired with bounding-box labels and class identifiers. In this work, annotations were generated using the open-source Computer Vision Annotation Tool (CVAT)[7].

The annotated dataset was randomly partitioned into training and validation subsets using an 80:20 split. Model optimization and performance assessment were carried out exclusively on the validation set, for which ground-truth annotations are available. The independent test dataset was not used at any stage during training or validation, and therefore does not admit standard quantitative performance metrics. It serves solely for post-training inference and candidate discovery.

[7] https://www.cvat.ai/

Model training and inference were executed on the Param Shakti high-performance computing (HPC) (Appendix B) facility at IIT Kharagpur. The use of HPC resources enabled efficient optimization of the high-capacity YOLOv11x–OBB model, facilitated experimentation with multiple configurations, and supported large-scale inference across the full rejected sample.

Figure A1 illustrates the internal training loop executed at each epoch. During each iteration:

(i) A batch of augmented images and corresponding OBB annotations is loaded.
(ii) A forward pass predicts bounding boxes, objectness scores, and class probabilities.
(iii) The composite detection loss $\mathcal{L}_{\text{total}}$ is computed.
(iv) A backward pass propagates gradients through the network.
(v) Model weights are updated via stochastic optimization to minimize detection error.

After each epoch, the model is evaluated on the validation set without gradient updates. Performance metrics (Precision, Recall, $\mathrm{F}_1$-score, mAP, and related diagnostics) are recorded. We summarize the evaluation metrics used for validation in the remainder of this section, as they are central to the interpretation of our results.

- **Precision ($P$):** The fraction of predicted nuclear detections that correspond to genuine nuclei in the validation set. High precision indicates effective suppression of false-positive detections arising from stars, host-galaxy substructure, or image artifacts.
- **Recall ($R$):** The fraction of ground-truth nuclear components successfully detected by the model. Recall quantifies completeness of nucleus recovery.
- **$\mathrm{F}_1$-score:** The harmonic mean of precision and recall, providing a balanced scalar measure of detection reliability and completeness.
- **Accuracy:** The fraction of correctly classified outcomes relative to all evaluated cases. Owing to intrinsic class imbalance, accuracy is reported for completeness but is not used as the primary performance indicator.
- **Positive Likelihood Ratio ($LR^+$):** Defined as the ratio of sensitivity to the false-positive rate, $LR^+$ quantifies how strongly a positive detection increases the likelihood that a predicted nucleus corresponds to a genuine nuclear component.
- **Mean Average Precision** (mAP): A threshold-independent summary of detection quality that combines localisation and classification performance. For a given Intersection-over-Union (IoU) matching threshold, a detection is counted as a true positive only if it is assigned the correct class *and* its predicted bounding box overlaps the ground-truth box with IoU above that threshold. The Average Precision (AP) is then the area under the resulting precision–recall curve, and the mAP is the mean of the AP values over the detection classes. We report two standard variants: $\mathrm{mAP}_{50}$, evaluated at a single IoU threshold of 0.50, and $\mathrm{mAP}_{50\text{-}95}$, averaged over ten IoU thresholds from 0.50 to 0.95 in steps of 0.05. The latter is the stricter, more localisation-sensitive metric.

The corresponding mathematical definitions are:

$$\text{Precision} = \frac{TP}{TP+FP}, \qquad \text{Recall} = \frac{TP}{TP+FN}, \tag{A9}$$

$$\mathrm{F}_1 = 2 \cdot \frac{\text{Precision} \cdot \text{Recall}}{\text{Precision} + \text{Recall}}, \tag{A10}$$

$$\text{Accuracy} = \frac{TP+TN}{TP+TN+FP+FN}, \tag{A11}$$

$$LR^+ = \frac{\text{Sensitivity}}{1-\text{Specificity}} = \frac{\dfrac{TP}{TP+FN}}{1-\dfrac{TN}{TN+FP}}. \tag{A12}$$

The mean Average Precision provides a single threshold-independent figure of merit that, unlike the pointwise metrics above, jointly rewards correct classification and accurate localisation across the full operating range of the detector. The Average Precision for a single class is the area under its precision–recall curve,

$$\mathrm{AP} = \int_0^1 P(R)\,\mathrm{d}R, \tag{A13}$$

where $P(R)$ is the precision as a function of recall, both evaluated at a fixed IoU matching threshold. The mean Average Precision averages this quantity over the $N_\mathrm{c}$ detection classes,

$$\mathrm{mAP} = \frac{1}{N_\mathrm{c}} \sum_{i=1}^{N_\mathrm{c}} \mathrm{AP}_i. \tag{A14}$$

Evaluating Equation (A14) at a single IoU threshold of 0.50 yields $\mathrm{mAP}_{50}$, while averaging it over the ten thresholds $\{0.50, 0.55, \ldots, 0.95\}$ yields $\mathrm{mAP}_{50\text{-}95}$:

$$\mathrm{mAP}_{50\text{-}95} = \frac{1}{10} \sum_{t\in\{0.50,\,0.55,\,\ldots,\,0.95\}} \mathrm{mAP}(t). \tag{A15}$$

Because $\mathrm{mAP}_{50\text{-}95}$ penalises imprecise bounding boxes more heavily, we adopt it as the primary monitoring metric for checkpoint selection and early stopping (Equations A17–A18), while $\mathrm{mAP}_{50}$ is reported alongside precision, recall, and F1 in the training-evolution diagnostics of Figure 3.

Here, $TP$ denotes correctly detected nuclear components, $FP$ denotes spurious detections not associated with any ground-truth nucleus, $FN$ denotes missed nuclear components, and $TN$ denotes correctly identified non-nuclear regions.

Model selection during training was guided primarily by the evolution of the validation $\mathrm{F}_1$-score, which provides a robust performance indicator under class imbalance and balances completeness against contamination.

To prevent overfitting and avoid unnecessary computation beyond convergence, the YOLO training framework employs an automatic early stopping mechanism, described formally below.

### *Early Stopping and Checkpoint Selection*

The early stopping mechanism is governed by a patience parameter $p \in \mathbb{Z}^+$. Let $\mathcal{M}(e)$ denote the primary monitoring metric evaluated on the validation set at epoch $e$, defined here as the mean Average Precision at IoU threshold 0.50, denoted $\mathrm{mAP}_{50}$. The best observed value of this metric up to and including epoch $e$ is tracked as:

$$\mathcal{M}^*(e) = \max_{e' \le e} \mathcal{M}(e'). \tag{A16}$$

A non-improvement counter $c(e)$ is maintained and updated at each epoch according to:

$$c(e) = \begin{cases} 0 & \text{if } \mathcal{M}(e) > \mathcal{M}^*(e-1), \\ c(e-1)+1 & \text{otherwise.} \end{cases} \tag{A17}$$

Training is terminated automatically when $c(e) \geq p$, that is, when no improvement in $\mathrm{mAP}_{50}$ is recorded over $p$ consecutive epochs. The default patience in the Ultralytics YOLOv11 implementation is $p = 50$ epochs. The optimal model checkpoint $e^*$ is defined independently of the stopping epoch as:

$$e^* = \arg\max_e \mathcal{M}(e), \tag{A18}$$

and the weights corresponding to $e^*$ are preserved automatically throughout training, ensuring that the final deployed model reflects peak validation performance rather than the state at termination.

In the baseline model, training terminated at epoch 293, with the optimal checkpoint identified at $e^* = 212$. The gap of 81 epochs between the best checkpoint and the stopping epoch arises directly from the patience window of $p = 50$ epochs: once $\mathrm{mAP}_{50}$ ceased improving at epoch 212, minor stochastic fluctuations in subsequent epochs periodically reset the counter before the full patience budget was exhausted, extending the run to epoch 293. The total training duration is therefore not a manually tuned hyperparameter but an emergent property of the convergence dynamics under the default patience schedule, consistent with the general empirical observation that models of this class reach stable optima within approximately 300 epochs. For the final YOLOv11x–OBB model, the same stopping mechanism was applied, with training terminating at epoch 237 and the optimal checkpoint selected at $e^* = 137$.

## APPENDIX B: PARAM SHAKTI SUPERCOMPUTING INFRASTRUCTURE

Training and validation of the final detection model were performed on the PARAM Shakti High-Performance Computing (HPC) facility at the Indian Institute of Technology Kharagpur. PARAM Shakti is a national-level academic supercomputing resource designed to support large-scale scientific computing and data-intensive workloads.

The YOLOv11x–OBB model, owing to its high parameter count and oriented bounding box regression, requires substantial GPU acceleration for efficient training. All experiments reported in this work were executed on GPU-enabled compute nodes equipped with **NVIDIA Tesla V100 GPUs**. The software environment was configured to support `CUDA 12.4`, ensuring compatibility with the Ultralytics YOLOv11 training framework and associated deep learning libraries.

Job scheduling and resource allocation were managed using the `SLURM` workload manager. A dedicated SLURM submission script was used to request computational resources, define memory and GPU requirements, and control the training job lifecycle. Model training was conducted using a single GPU node, which was sufficient given the dataset size and batch configuration adopted in this study.

The total wall-clock time required to complete training, including validation and checkpointing, was approximately **6.0 hours**. This configuration provided an effective balance between computational efficiency and training stability, allowing multiple experimental iterations during model development without excessive resource overhead.

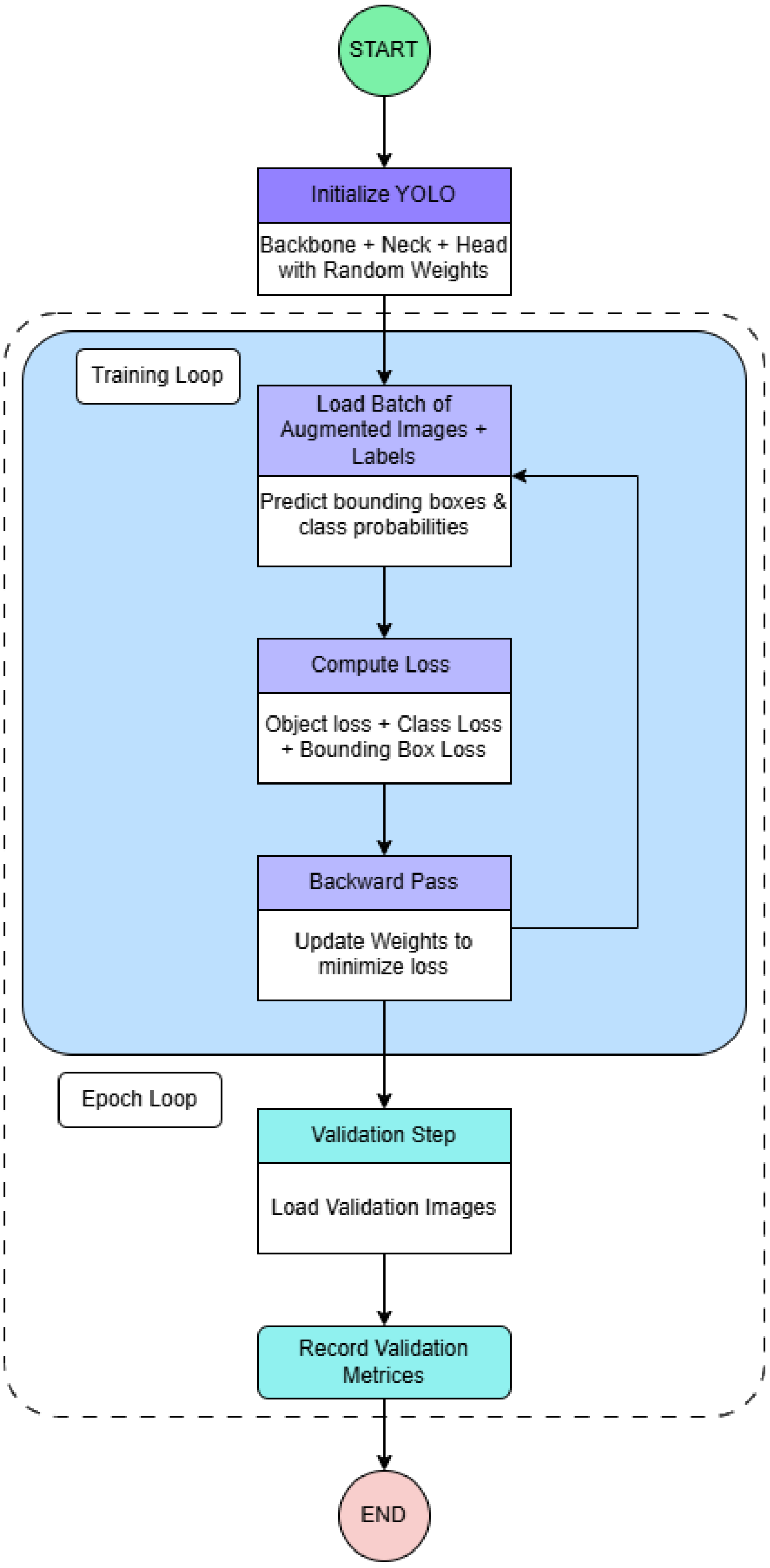

**Figure A1.** Schematic representation of the YOLO training and validation cycle. Each epoch consists of repeated forward and backward passes over augmented mini-batches, followed by validation evaluation.

The Python training script executed within the SLURM job invoked the Ultralytics YOLO training pipeline for the `YOLOv11x-obb` architecture, using the dataset configuration and hyperparameters described in Appendix A. The overall training and inference workflow is schematically illustrated in Figure 1.

## APPENDIX C: OVERLAPPING CLASSES FILTRATION ALGORITHM

To suppress ambiguous detections arising from spatial overlap between predicted dual nuclei and foreground stars, we applied a geometry-based filtering stage to the model outputs. This step was designed to remove cases where class ambiguity is intrinsic to the image geometry and cannot be reliably resolved using photometric information alone.

Each prediction generated by the YOLOv11x-OBB model is stored in the oriented bounding box (OBB) format, where detections are represented by a class label followed by four normalized corner coordinates:

$$[\texttt{class}, x_1, y_1, x_2, y_2, x_3, y_3, x_4, y_4], \tag{C1}$$

with $(x_i, y_i) \in [0, 1]$. These coordinates define a closed polygon in image space corresponding to the predicted object extent.

For every image, all predicted OBBs were converted into polygonal geometries using the `Shapely` computational geometry library. Pairwise intersection tests were then performed between all detections belonging to different classes. Specifically, if a detection labeled as a dual nucleus (Class 0) spatially intersected with a detection labeled as a foreground star (Class 1), the intersection area was computed. An image was flagged for exclusion if the intersection area satisfied

$$\mathrm{Area}(\mathcal{P}_0 \cap \mathcal{P}_1) > 0, \tag{C2}$$

where $\mathcal{P}_0$ and $\mathcal{P}_1$ denote the OBB polygons corresponding to Class 0 and Class 1 detections, respectively.

This criterion ensures that only genuine geometric overlaps rather than simple proximity trigger removal. All flagged detections were logged automatically and exported as a structured CSV file containing the corresponding image identifiers. This procedure was applied uniformly across the full detected dataset without any manual intervention, ensuring reproducibility and eliminating human bias.

By enforcing geometric consistency between class predictions, this filtering step significantly reduces residual false positives while preserving isolated dual-nucleus detections and well-separated merger candidates.

## APPENDIX D: PROJECTED SEPARATION ESTIMATION ALGORITHM

A key objective of this study is to quantify the minimum projected physical separation between detected dual-nucleus systems. Direct estimation of three-dimensional separations is not feasible for the majority of detections due to the absence of spectroscopic redshifts for both components. We therefore adopt a physically motivated approximation to estimate projected separations using image-plane geometry combined with available spectroscopic information.

For each post-filtered detection labelled as a dual nucleus (Class 0), the YOLOv11x-OBB model outputs an oriented bounding box (OBB) enclosing both nuclear components. The OBB is represented by four normalised corner coordinates in image space:

$$(x_1, y_1),\ (x_2, y_2),\ (x_3, y_3),\ (x_4, y_4), \tag{D1}$$

with all coordinates scaled to the unit interval. These coordinates were first converted to pixel space assuming a fixed image size of $120 \times 120$ pixels.

The projected angular separation was estimated by computing the Euclidean lengths of the four edges of the OBB in pixel space and selecting the maximum edge length:

$$\theta_{\rm px} = \max_{i \in \{1,\dots,4\}} \|\mathbf{p}_i - \mathbf{p}_{i+1}\|, \tag{D2}$$

where $\mathbf{p}_i$ denotes the pixel coordinates of the $i$-th corner and indices are taken cyclically. This choice provides a conservative upper bound on the projected separation between the two nuclei, ensuring that both components are fully enclosed within the estimated distance.

The angular separation in arcseconds was then obtained using the SDSS image scale of 0.3 arcsec pixel$^{-1}$:

$$\theta_{arcsec} = 0.3 \times \theta_{\rm px}. \tag{D3}$$

To convert angular separations to physical distances, we assume that both nuclei reside at the same redshift as the primary galaxy nucleus, whose spectroscopic redshift $z$ is available from the SDSS catalogue. The projected physical separation follows from $s = \theta_{\rm rad}\, D_A(z)$, the product of the angular separation in radians and the angular diameter distance. We adopt the low-redshift (linear Hubble-law) approximation to the angular diameter distance,

$$D_A(z) \approx \frac{c\,z}{H_0}, \tag{D4}$$

which neglects the $\Omega_m$-dependent integral and the $(1+z)$ term of the exact $\Lambda$CDM expression. Converting $\theta_{arcsec}$ from arcsec to radians and the resulting distance from Mpc to kpc, the projected physical separation in kpc is

$$s_{\rm kpc} = \frac{\pi \times 10^3\, c}{180 \times 3600 \times H_0}\, \theta_{arcsec}\, z, \tag{D5}$$

with $c = 3 \times 10^5$ km s$^{-1}$ and $H_0 = 71$ km s$^{-1}$ Mpc$^{-1}$, consistent with the cosmology adopted throughout this work ($\Omega_m = 0.3$).

Above equation is accurate in the low-redshift regime but progressively overestimates $D_A(z)$ relative to the exact $\Lambda$CDM value as redshift increases, by of order a few per cent at $z \approx 0.1$ and rising to tens of per cent by $z \approx 0.3$. The absolute physical separations reported here should be interpreted with this in mind; the dimensionless cross-pipeline ratio of Appendix F is unaffected, since the conversion factor cancels.

## APPENDIX E: CONFIDENCE INTERVAL ESTIMATION FOR SENSITIVITY ANALYSIS

To quantify the uncertainty in the manually estimated fraction of bona fide dual-nucleus systems, we model the classification outcome as a binomial process. Let $X \sim \mathrm{Binomial}(n, p)$ denote the number of true dual-nucleus systems identified in a sample of size $n$, where $p$ is the underlying probability that a randomly selected detection corresponds to a genuine dual nucleus. The empirical estimator for this probability is:

$$\hat{p} = \frac{k}{n}, \tag{E1}$$

where $k$ is the number of detections classified as dual nuclei. For sufficiently large $n$, the sampling distribution of $\hat{p}$ can be approximated as normal with variance:

$$\mathrm{Var}(\hat{p}) = \frac{\hat{p}(1-\hat{p})}{n}. \tag{E2}$$

The corresponding standard error is:

$$\sigma_{\hat{p}} = \sqrt{\frac{\hat{p}(1-\hat{p})}{n}}. \tag{E3}$$

A two-sided $(1-\alpha)$ confidence interval is then given by:

$$\hat{p} \pm z_{1-\alpha/2}\, \sigma_{\hat{p}}, \tag{E4}$$

where $z_{1-\alpha/2}$ is the critical value of the standard normal distribution. For a 95% confidence level, $z_{0.975} \approx 1.96$. Applying this formulation to the three independent samples:

- **Full sample ($n = 200$, $k = 109$):**

$$\hat{p} = 0.545, \tag{E5}$$

$$\sigma_{\hat{p}} \approx \sqrt{\frac{0.545 \times 0.455}{200}} \approx 0.035, \tag{E6}$$

yielding a 95% confidence interval of $[0.476, 0.614]$.

- **Compact-separation subset ($d \leq 6.87''$) ($n = 100$, $k = 59$):**

$$\hat{p} = 0.59, \tag{E7}$$

$$\sigma_{\hat{p}} \approx \sqrt{\frac{0.59 \times 0.41}{100}} \approx 0.049, \tag{E8}$$

yielding a 95% confidence interval of $[0.494, 0.686]$.

- **Cumulative $\leq 9.79''$ subset ($n = 100$, $k = 62$):**

$$\hat{p} = 0.62, \tag{E9}$$

$$\sigma_{\hat{p}} \approx \sqrt{\frac{0.62 \times 0.38}{100}} \approx 0.048, \tag{E10}$$

yielding a 95% confidence interval of $[0.525, 0.715]$.

These intervals are used in Section 4.5 to assess statistical consistency across subsets and to propagate uncertainty into population-level estimates.

## APPENDIX F: CROSS-PIPELINE CALIBRATION OF PROJECTED SEPARATIONS

The conversion from the YOLO OBB edge length to a true centroid-to-centroid separation, used in Sections 4.3 and 4.4, is derived from a dedicated method-comparison analysis between the two independent pipelines applied to this sample.

### F1 Data and rationale

The calibration requires systems for which a projected separation is available from *both* pipelines. Such systems are necessarily restricted to the GOTHIC-confirmed dual-nucleus catalogue of Bhattacharya et al. (2023), because GOTHIC reports a centroid-to-centroid separation only for the sources it retains. The rejected sample analysed in this work carries no GOTHIC separation by construction. We therefore draw the calibration set from the held-out validation split of the final model (Table 1), which is composed of these GOTHIC-derived dual-nucleus systems together with foreground stars and is disjoint from the training data. Using the validation split rather than the training images ensures that the calibration is unaffected by any fit memorised during training.

Of the 624 validation images, 221 contain a dual-nucleus system that YOLO detected and for which GOTHIC provides an independent centroid-to-centroid separation. These 221 common systems constitute the calibration sample. For each, the separation is available as two measurements expressed in identical units: an angular pair — the YOLO OBB edge length $B$ and the GOTHIC centroid distance $D$, both in arcsec, and a physical pair $C$ and $E$ in kpc, where the kpc values are obtained from the same angular measurements via the angular-diameter distance of Appendix D. Both $B$ and $D$ are noisy measurements of the same on-sky quantity, and neither is a controlled predictor of the other; establishing the mapping between them is therefore a method-comparison problem rather than a predictor–response regression.

**Table F1.** Slope estimates for each column pair ($n = 221$). Deming regression assumes $\lambda = 1$ (orthogonal regression). The two OLS fits bracket the symmetric estimators; their separation reflects the attenuation bias arising from measurement error in both variables.

| Estimator | Slope | Intercept | Role |
|---|---|---|---|
| | *B–D (arcsec)* | | |
| Deming / orthogonal | 1.1698 | −4.8283 | Recommended relationship |
| Reduced major axis | 1.1531 | −4.6603 | Relationship estimate |
| OLS $D \mid B$ | 1.0464 | −3.5893 | Prediction only (attenuated) |
| OLS $B \mid D$ (inv.) | 1.2706 | −5.8405 | Bracketing line |
| | *C–E (kpc)* | | |
| Deming / orthogonal | 0.7738 | −2.0516 | Recommended relationship |
| Reduced major axis | 0.7928 | −2.4376 | Relationship estimate |
| OLS $E \mid C$ | 0.7163 | −0.8853 | Prediction only (attenuated) |
| OLS $C \mid E$ (inv.) | 0.8775 | −4.1556 | Bracketing line |

### F2 Why not ordinary least squares

Ordinary least squares (OLS) of $D$ on $B$ minimises only the vertical residuals and assumes the predictor is error-free. That assumption fails here, with two consequences. First, error in the predictor biases the OLS slope toward zero (regression dilution): the OLS angular slope is 1.05, whereas an estimator that accounts for error in both variables gives 1.17. The difference is bias, not signal. Second, regressing $D$ on $B$ and $B$ on $D$ yields different lines that bracket the true relation rather than agreeing on it, so the result depends on the arbitrary choice of predictor. We therefore use a symmetric, errors-in-variables estimator for the relationship and Bland–Altman analysis for the agreement, retaining OLS only as a one-directional YOLO→GOTHIC prediction.

### F3 Errors-in-variables regression

We adopt Deming (orthogonal) regression with error-variance ratio $\lambda = \mathrm{var}(\varepsilon_x)/\mathrm{var}(\varepsilon_y) = 1$, cross-checked against reduced major axis (RMA). Table F1 collects the slope estimates; the two OLS rows bracket the symmetric estimates, and their separation is the attenuation bias.

For the angular pair (Pearson $r = 0.907$) the recommended relationship is $D \approx 1.17\,B - 4.83\,''$, with the true slope bracketed in $[1.05,\ 1.27]$; the OLS value sits at the low end purely through attenuation. For the physical pair (Pearson $r = 0.903$) Deming gives $E \approx 0.77\,C - 2.05$, bracket $[0.72,\ 0.88]$. As shown below, the physical relation is consistent with a purely multiplicative form, so the kpc intercept is not physically meaningful.

### F4 Bland–Altman agreement

Bland–Altman analysis plots the difference between the two methods against their mean. The appropriate form differs by pair. For the angular pair the difference is additive: GOTHIC lies on average $3.12\,''$ below YOLO, with 95% limits of agreement $[-5.81,\ -0.44]\,''$. The difference is not constant but shrinks as separation grows (proportional bias, slope $+0.149$, $p \approx 10^{-6}$). An additive angular gap that matters most, in relative terms, for close pairs is the signature of YOLO measuring a detection-box/edge-type separation while GOTHIC measures the centroid-to-centroid distance.

For the physical pair the relation is multiplicative and is better summarised by the ratio: $E/C$ has mean 0.663 with limits $[0.32,\ 1.01]$, essentially flat across the separation range.

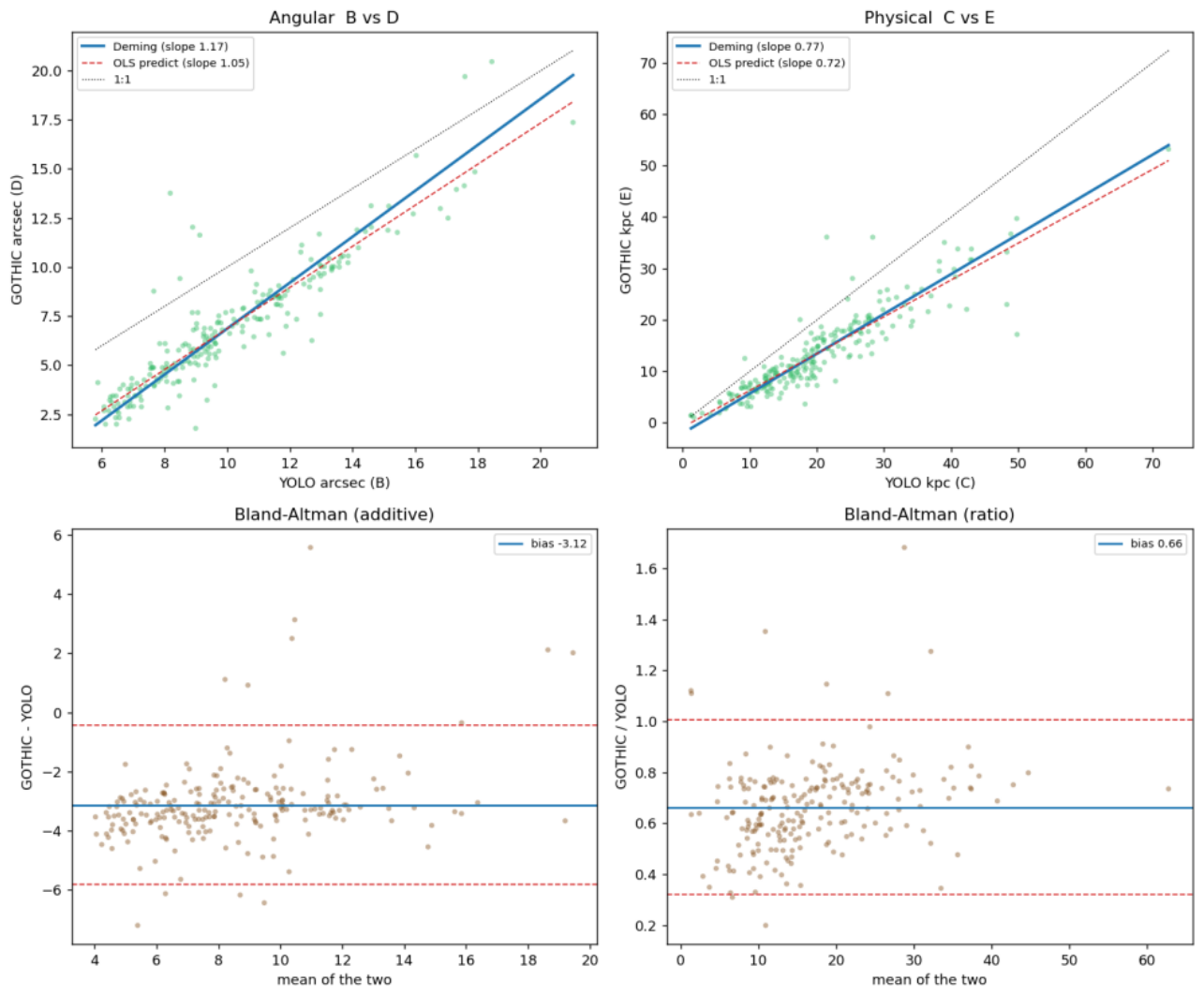

**Figure F1.** Top: GOTHIC versus YOLO with the Deming relationship (solid), the OLS prediction line (dashed) and the 1:1 line (dotted); the Deming line sits above OLS, visualising the attenuation. Bottom: Bland–Altman agreement — additive for the angular pair (left) and ratio for the physical pair (right), with the bias (solid) and 95% limits of agreement (dashed).

## F5 Adopted calibration and unit invariance

We adopt the angular Deming relation, Equation 2, as the operational YOLO→GOTHIC conversion, because the angular relationship genuinely has both a slope $\neq 1$ and a real offset, and Deming is the unbiased symmetric estimator. Physical separations are then obtained by applying the *same* per-object angular-diameter scale $k(z)$ (Appendix D) to the corrected angular value.

Because both pipelines' angular measurements are mapped to physical units with this same per-object scale, the cross-pipeline ratio is identical in arcsec and kpc:

$$\frac{D_i}{B_i} = \frac{E_i}{C_i} = r_i, \qquad \max_i \left| \frac{D_i}{B_i} - \frac{E_i}{C_i} \right| = 4.4 \times 10^{-16}, \tag{F1}$$

with $\bar{r} = 0.663$ and $\sigma_r = 0.175$. We stress that this equality is a property of the shared unit conversion rather than independent evidence of agreement in distance: it follows by construction because $C = k(z)\,B$ and $E = k(z)\,D$ with the same $k(z)$, which is why the residual is at the level of floating-point precision. The single number $r = 0.66 \pm 0.17$ is a convenient coarse descriptor ("GOTHIC is on average $\sim 2/3$ of YOLO"), but it is not a constant: combining it with Equation 2 gives $r_i = D_i/B_i = 1.17 - 4.83/B_i$, which increases monotonically with separation. The scatter $\sigma_r = 0.175$ is therefore dominated by this deterministic separation dependence rather than by random measurement error, and the affine relation of Equation 2 is the physically faithful description.

This paper has been typeset from a TEX/LATEX file prepared by the author.

# Supplementary Material to "Decoupling candidate dual AGN from chance superpositions in the GOTHIC survey via a deep-learning framework"

**Table 1.** Sub-kiloparsec dual-nucleus candidates.

| No. | objID | RA (deg) | Dec (deg) | Redshift ($z$) |
|---|---|---|---|---|
| 1 | 1237664854177808662 | 206.97879 | 36.18467 | 0.00783653 |
| 2 | 1237657775541125383 | 130.65808 | 37.59502 | 0.00676495 |
| 3 | 1237652899689070767 | 34.08085 | -9.47011 | 0.00671386 |
| 4 | 1237655109455708383 | 203.40578 | 59.71563 | 0.00967897 |
| 5 | 1237662225161584991 | 239.00703 | 26.28398 | 0.0087119 |
| 6 | 1237658424624611527 | 153.20469 | 7.10375 | 0.00515148 |
| 7 | 1237653436544712833 | 346.48445 | -10.04914 | 0.00760355 |
| 8 | 1237655464302870669 | 171.97940 | 61.62648 | 0.0076261 |
| 9 | 1237661813348303000 | 184.36687 | 12.93234 | 0.00686889 |
| 10 | 1237665566068506761 | 210.57453 | 22.72651 | 0.0103324 |
| 11 | 1237667253477245004 | 180.60503 | 29.47034 | 0.0112294 |
| 12 | 1237661959371948039 | 213.78837 | 49.34228 | 0.0130732 |
| 13 | 1237659343736406092 | 188.58931 | 8.24049 | 0.0061483 |
| 14 | 1237655369279799401 | 148.72947 | 56.60778 | 0.0053138 |
| 15 | 1237657606427181270 | 131.48077 | 41.52168 | 0.00926618 |
| 16 | 1237652629101936941 | 7.86481 | -10.67583 | 0.0117493 |
| 17 | 1237668590260125784 | 185.48279 | 17.68113 | 0.00675247 |
| 18 | 1237661125075534003 | 126.64071 | 25.49979 | 0.00716572 |
| 19 | 1237661852538109966 | 168.54787 | 44.96812 | 0.00736419 |
| 20 | 1237661418748706906 | 210.96849 | 55.59233 | 0.00656888 |
| 21 | 1237662525231005777 | 189.76960 | 14.73113 | 0.00415325 |
| 22 | 1237668624622551206 | 183.70734 | 17.71898 | 0.00791893 |
| 23 | 1237655368744632621 | 155.66636 | 58.09002 | 0.00675756 |
| 24 | 1237661849865093393 | 204.64433 | 42.26256 | 0.008221 |
| 25 | 1237651737907167300 | 186.17210 | 3.72146 | 0.00620117 |
| 26 | 1237656971302010958 | 12.99888 | -0.48928 | 0.00538411 |
| 27 | 1237663784216101143 | 42.71672 | 0.00339 | 0.00529295 |
| 28 | 1237653612110938117 | 140.23300 | 52.56871 | 0.00766776 |
| 29 | 1237652935640679317 | 310.48710 | -5.29904 | 0.0124006 |
| 30 | 1237664667367047289 | 171.01121 | 38.21979 | 0.00740596 |
| 31 | 1237655126083567750 | 185.82685 | 5.60716 | 0.00601899 |
| 32 | 1237667254007103520 | 162.30333 | 27.92259 | 0.0047693 |
| 33 | 1237658423021469814 | 170.34761 | 6.66712 | 0.00889607 |
| 34 | 1237665367972773964 | 174.75510 | 31.48765 | 0.00905005 |
| 35 | 1237667911669383440 | 191.10396 | 24.90372 | 0.00609563 |
| 36 | 1237671129667666324 | 181.19623 | 10.62665 | 0.00851305 |
| 37 | 1237651192434655452 | 138.56913 | 57.04434 | 0.0083881 |
| 38 | 1237662306197176380 | 234.08101 | 30.68232 | 0.00585922 |
| 39 | 1237655108374429953 | 169.22468 | 59.13278 | 0.00561024 |
| 40 | 1237655124466598012 | 171.27261 | 4.12100 | 0.00535795 |
| 41 | 1237661874021072953 | 207.09733 | 43.70848 | 0.00762431 |
| 42 | 1237661149771858254 | 208.69553 | 46.33476 | 0.00561195 |
| 43 | 1237658297917767914 | 145.57891 | 4.68939 | 0.00666106 |
| 44 | 1237648705134264658 | 229.73769 | 0.51569 | 0.00694762 |
| 45 | 1237654342253674785 | 210.53624 | 61.41240 | 0.00564085 |

## THE SUB-KILOPARSEC CANDIDATE SAMPLE

This supplement lists and illustrates the most compact dual-nucleus candidates identified in this work: the 212 YOLOv11x–OBB detections whose GOTHIC-calibrated centroid-to-centroid separation satisfies $s_{\rm GOTHIC} \leq 1$ kpc, i.e. the leftmost bin of the projected physical-separation distribution (Section 4.7 of the main paper). At sub-kiloparsec separations the two nuclei are unresolved in SDSS imaging, so these are dual-nucleus *candidates* rather than confirmed dual AGN. Table 1 gives their SDSS `objID`, coordinates and redshift, and Fig. 1 shows the corresponding image cutouts; each `objID` in the table hyperlinks to its cutout.

**Table 1.** Sub-kiloparsec dual-nucleus candidates (Table 1, continued).

| No. | objID | RA (deg) | Dec (deg) | Redshift ($z$) |
|---|---|---|---|---|
| 46 | 1237655464310472782 | 207.72398 | 60.14174 | 0.00792412 |
| 47 | 1237664667903918385 | 170.94341 | 38.58240 | 0.00524803 |
| 48 | 1237661970648531070 | 184.66095 | 6.70838 | 0.00668592 |
| 49 | 1237661434855358615 | 217.45539 | 42.93971 | 0.00827967 |
| 50 | 1237664668959048025 | 124.44168 | 22.67806 | 0.00715066 |
| 51 | 1237658302207623377 | 170.65736 | 58.32852 | 0.0064684 |
| 52 | 1237648720698867902 | 191.44958 | -0.43224 | 0.00555998 |
| 53 | 1237667321648971805 | 191.54482 | 26.25023 | 0.00630545 |
| 54 | 1237668272976625726 | 213.14036 | 18.50036 | 0.00738297 |
| 55 | 1237651821637337318 | 219.81433 | 3.27265 | 0.00528603 |
| 56 | 1237660765912957126 | 132.60499 | 32.62190 | 0.00740937 |
| 57 | 1237661813879341072 | 170.57733 | 13.06505 | 0.00524275 |
| 58 | 1237658304886079711 | 146.59535 | 54.86899 | 0.00529389 |
| 59 | 1237658613593669710 | 176.27606 | 50.30068 | 0.00557602 |
| 60 | 1237658204497510545 | 140.80656 | 40.46567 | 0.0057912 |
| 61 | 1237662305112031303 | 204.09524 | 39.70460 | 0.00819632 |
| 62 | 1237655691404181935 | 224.08465 | -2.76163 | 0.00629797 |
| 63 | 1237655464304050383 | 177.36083 | 62.05572 | 0.00431745 |
| 64 | 1237671266570993942 | 191.14330 | 1.27530 | 0.00469413 |
| 65 | 1237662263780704489 | 216.74907 | 8.68362 | 0.0045753 |
| 66 | 1237668292299587626 | 170.00278 | 18.26050 | 0.00545912 |
| 67 | 1237653612113166459 | 147.37626 | 55.57971 | 0.00525211 |
| 68 | 1237667255621058660 | 170.77901 | 30.47892 | 0.00536102 |
| 69 | 1237662636367872012 | 222.59431 | 11.40391 | 0.00593264 |
| 70 | 1237648704055673107 | 218.72583 | -0.34265 | 0.0060044 |
| 71 | 1237671692301893932 | 245.31777 | 18.61831 | 0.00812646 |
| 72 | 1237671127517626432 | 179.75312 | 4.66964 | 0.00536336 |
| 73 | 1237655107301277861 | 171.85620 | 58.63879 | 0.00418212 |
| 74 | 1237655125020049616 | 209.30851 | 4.30732 | 0.00404988 |
| 75 | 1237670956792348829 | 33.92156 | -8.50807 | 0.00462104 |
| 76 | 1237661358620082187 | 191.59724 | 48.23519 | 0.00300073 |
| 77 | 1237661416603975998 | 219.42217 | 51.45734 | 0.00692239 |
| 78 | 1237660412115288426 | 143.09234 | 7.30229 | 0.007109 |
| 79 | 1237665533333471631 | 240.05761 | 17.84830 | 0.00684895 |
| 80 | 1237656495113765145 | 0.26499 | 14.58019 | 0.00571765 |
| 81 | 1237650369408270425 | 180.19764 | -3.42011 | 0.00497801 |
| 82 | 1237662195076038866 | 221.93561 | 36.50465 | 0.00408862 |
| 83 | 1237667485916201224 | 131.35586 | 15.32946 | 0.00538601 |
| 84 | 1237655126093594767 | 208.76879 | 5.18949 | 0.00468982 |
| 85 | 1237658800961749081 | 175.27821 | 53.79782 | 0.00440259 |
| 86 | 1237671127518543970 | 180.76172 | 6.50132 | 0.00437887 |
| 87 | 1237664854190129505 | 237.79070 | 25.82555 | 0.00724367 |
| 88 | 1237665096854863891 | 147.54659 | 28.01245 | 0.00481019 |
| 89 | 1237654606947156176 | 184.84924 | 6.23302 | 0.00623633 |
| 90 | 1237668298745970807 | 197.11838 | 20.03387 | 0.00495315 |
| 91 | 1237668290698739843 | 165.77754 | 20.06019 | 0.00429376 |
| 92 | 1237664869750669476 | 143.67397 | 32.53432 | 0.00463259 |
| 93 | 1237661382234472578 | 143.42943 | 33.60030 | 0.00522282 |
| 94 | 1237655369820471426 | 163.62375 | 60.72413 | 0.00440912 |
| 95 | 1237654879129764079 | 222.59524 | 2.95854 | 0.00566138 |
| 96 | 1237651271359725676 | 163.28006 | 64.79530 | 0.00367905 |
| 97 | 1237661976547623019 | 186.02124 | 8.29383 | 0.00415325 |
| 98 | 1237658611446513886 | 177.73382 | 48.53160 | 0.0037607 |
| 99 | 1237652900762157127 | 32.49889 | -8.83643 | 0.00522346 |
| 100 | 1237661849868763416 | 215.22799 | 39.91419 | 0.00558964 |
| 101 | 1237661382774489386 | 151.12261 | 36.94796 | 0.00522126 |
| 102 | 1237652901303812119 | 43.44460 | -7.39555 | 0.00449437 |
| 103 | 1237650795146182725 | 146.00779 | -0.64227 | 0.0047763 |
| 104 | 1237662236930539554 | 187.22900 | 9.42111 | 0.00348355 |
| 105 | 1237658492283781293 | 190.25445 | 9.71845 | 0.00453878 |
| 106 | 1237663916810043683 | 139.39957 | 65.37632 | 0.00558786 |

**Table 1.** Sub-kiloparsec dual-nucleus candidates (Table 1, continued).

| No. | objID | RA (deg) | Dec (deg) | Redshift ($z$) |
|---|---|---|---|---|
| 107 | 1237664669510402089 | 158.84467 | 37.67166 | 0.00567336 |
| 108 | 1237667254540566807 | 153.76810 | 26.56742 | 0.00458432 |
| 109 | 1237662224594763892 | 157.43077 | 36.27162 | 0.00390106 |
| 110 | 1237661851467579519 | 178.96587 | 45.16267 | 0.00349629 |
| 111 | 1237655126093594752 | 208.76127 | 5.09031 | 0.00465898 |
| 112 | 1237667254002253912 | 150.22818 | 25.23136 | 0.00479211 |
| 113 | 1237668292833574973 | 163.08905 | 17.93527 | 0.00434835 |
| 114 | 1237673705579872326 | 120.09922 | 42.19380 | 0.00240752 |
| 115 | 1237662239077892488 | 186.93452 | 11.21456 | 0.00475753 |
| 116 | 1237662306725462159 | 212.50287 | 39.19836 | 0.00495963 |
| 117 | 1237658629159977157 | 190.61045 | 11.74030 | 0.00375927 |
| 118 | 1237667322709016589 | 156.92304 | 24.27155 | 0.00411442 |
| 119 | 1237648720694476862 | 181.33292 | -0.48020 | 0.00457376 |
| 120 | 1237662239077630310 | 186.38111 | 11.15839 | 0.00295195 |
| 121 | 1237650761852780557 | 177.88897 | -2.37278 | 0.00341584 |
| 122 | 1237657590322757751 | 182.38225 | 54.93836 | 0.00331829 |
| 123 | 1237667550881710334 | 164.01250 | 23.81348 | 0.00406482 |
| 124 | 1237658491730592014 | 152.58550 | 7.75378 | 0.00427012 |
| 125 | 1237667323258994717 | 190.20705 | 27.56404 | 0.00372799 |
| 126 | 1237662194524881016 | 179.87147 | 42.34921 | 0.00295256 |
| 127 | 1237661358616150192 | 178.13934 | 48.29307 | 0.00350108 |
| 128 | 1237661950256087121 | 188.52808 | 12.74160 | 0.00503009 |
| 129 | 1237660634391969967 | 179.70702 | 46.46474 | 0.00264106 |
| 130 | 1237657857146748985 | 179.78967 | 52.70761 | 0.00360642 |
| 131 | 1237652900766351527 | 42.16665 | -7.81342 | 0.00483179 |
| 132 | 1237665023834783952 | 194.41897 | 33.69423 | 0.00293196 |
| 133 | 1237663916269633621 | 128.65495 | 59.09333 | 0.00480354 |
| 134 | 1237665227842388053 | 185.10746 | 33.24215 | 0.00349119 |
| 135 | 1237651736300617974 | 195.47220 | 2.46061 | 0.0029551 |
| 136 | 1237661966360969460 | 199.00244 | 40.96939 | 0.00375171 |
| 137 | 1237661355928060093 | 166.43474 | 44.83283 | 0.00262658 |
| 138 | 1237667212140347526 | 185.84046 | 30.56207 | 0.0022458 |
| 139 | 1237668293910397163 | 170.41804 | 19.61203 | 0.00383223 |
| 140 | 1237663531326767420 | 112.74540 | 41.16661 | 0.00297312 |
| 141 | 1237658800426975262 | 183.51035 | 53.75484 | 0.00305526 |
| 142 | 1237664338785533999 | 164.44584 | 36.26077 | 0.00202755 |
| 143 | 1237651754534568238 | 151.10472 | 2.55861 | 0.00375556 |
| 144 | 1237656494577942834 | 2.78584 | 14.23967 | 0.00267277 |
| 145 | 1237662302436196417 | 219.08101 | 47.56517 | 0.0024477 |
| 146 | 1237661873467359356 | 154.06515 | 41.16638 | 0.00172626 |
| 147 | 1237652944250077292 | 2.47036 | 15.73530 | 0.00287797 |
| 148 | 1237657856072351777 | 177.47687 | 51.73640 | 0.00313227 |
| 149 | 1237658630232604854 | 188.00318 | 12.62031 | 0.00164376 |
| 150 | 1237661064954118186 | 151.82776 | 10.36857 | 0.00182728 |
| 151 | 1237661068189565080 | 185.16784 | 13.88956 | 0.00266545 |
| 152 | 1237665129079832842 | 182.25660 | 35.59466 | 0.00240206 |
| 153 | 1237664836472144337 | 147.58119 | 31.45618 | 0.00178564 |
| 154 | 1237657628453699841 | 178.46819 | 51.49378 | 0.00149186 |
| 155 | 1237654386803278005 | 122.40582 | 41.59072 | 0.00231939 |
| 156 | 1237661357535592532 | 156.46165 | 43.93283 | 0.00218612 |
| 157 | 1237665129067905298 | 150.15226 | 30.53605 | 0.00162153 |
| 158 | 1237651274048602299 | 188.65992 | 68.34039 | 0.00153428 |
| 159 | 1237660765905748408 | 117.90017 | 22.58358 | 0.000564882 |

**Table 1.** Sub-kiloparsec dual-nucleus candidates (Table 1, continued).

| No. | objID | RA (deg) | Dec (deg) | Redshift ($z$) |
|---|---|---|---|---|
| 160 | 1237649929171632403 | 61.02435 | -4.99108 | 0.000566508 |
| 161 | 1237658492266348708 | 149.92908 | 7.87099 | 0.000264883 |
| 162 | 1237667781760319698 | 183.32949 | 21.74410 | 0.000239688 |
| 163 | 1237667430626033791 | 129.99837 | 16.86267 | 0.000207192 |
| 164 | 1237663783654326780 | 345.80129 | -0.38320 | 0.000350801 |
| 165 | 1237660765374447860 | 129.39015 | 30.01556 | 0.000233354 |
| 166 | 1237665126919700656 | 148.94642 | 28.38125 | 0.000337004 |
| 167 | 1237657774997963036 | 118.58910 | 27.46225 | 0.000254276 |
| 168 | 1237651752923168983 | 149.35646 | 1.34514 | 0.000195835 |
| 169 | 1237652598487122948 | 311.36363 | -7.01417 | 0.000221583 |
| 170 | 1237667536383770662 | 121.13799 | 10.26213 | 0.000180927 |
| 171 | 1237667911669842224 | 192.12358 | 24.96447 | 0.000220244 |
| 172 | 1237667448342118581 | 189.23540 | 25.69186 | 0.000215019 |
| 173 | 1237663458316255528 | 323.79183 | 0.56194 | 0.000108063 |
| 174 | 1237662224624845166 | 239.06879 | 25.82424 | 0.000122431 |
| 175 | 1237678595932094670 | 335.04954 | 1.14617 | 0.000136979 |
| 176 | 1237651497365012901 | 113.15860 | 39.32030 | 0.000150769 |
| 177 | 1237670964317126835 | 152.58138 | 16.41057 | 0.000161207 |
| 178 | 1237663786885316826 | 128.90379 | 58.92243 | 0.000153769 |
| 179 | 1237652901830066274 | 19.00887 | -8.84163 | 0.000140169 |
| 180 | 1237668584893251655 | 162.48483 | 15.70840 | 0.000110508 |
| 181 | 1237667783923531788 | 221.66252 | 19.28362 | 0.000128504 |
| 182 | 1237663787950211670 | 111.48923 | 41.71840 | 0.000105321 |
| 183 | 1237660413187129505 | 138.67955 | 7.59444 | 8.91645e-05 |
| 184 | 1237670458048054247 | 49.14367 | 41.04924 | 7.3561e-05 |
| 185 | 1237671939260088429 | 226.23761 | 56.41592 | 7.5489e-05 |
| 186 | 1237648704578453520 | 186.48049 | 0.20025 | 6.6139e-05 |
| 187 | 1237658492807676168 | 160.21403 | 9.15020 | 7.60792e-05 |
| 188 | 1237657192514978333 | 358.07077 | 1.12098 | 9.91854e-05 |
| 189 | 1237663457239630524 | 317.27499 | -0.34690 | 0.000135879 |
| 190 | 1237668367464923337 | 247.54376 | 11.68027 | 7.63871e-05 |
| 191 | 1237648720693821551 | 179.94674 | -0.47435 | 5.35782e-05 |
| 192 | 1237648720156098698 | 177.89586 | -0.94887 | 9.04981e-05 |
| 193 | 1237651272962080975 | 130.58902 | 54.83422 | 9.72971e-05 |
| 194 | 1237660961861665021 | 129.05854 | 29.05421 | 4.82239e-05 |
| 195 | 1237678536880947963 | 320.25435 | 12.01060 | 3.56939e-05 |
| 196 | 1237663542612197888 | 337.06190 | -0.43361 | 7.11645e-05 |
| 197 | 1237663529723363505 | 127.30266 | 54.49632 | 4.07283e-05 |
| 198 | 1237674290757238899 | 122.55695 | 35.47914 | 3.0678e-05 |
| 199 | 1237666302165713097 | 49.25939 | 1.25336 | 2.77344e-05 |
| 200 | 1237662524158443584 | 192.52648 | 13.81235 | 2.0593e-05 |
| 201 | 1237662223543042171 | 219.66589 | 33.39131 | 2.70439e-05 |
| 202 | 1237679454920376462 | 348.75548 | 15.66020 | 2.32449e-05 |
| 203 | 1237661353778282639 | 173.12835 | 57.52034 | 2.12021e-05 |
| 204 | 1237663542065037920 | 313.49661 | -0.92013 | 2.29874e-05 |
| 205 | 1237667142857261376 | 118.68560 | 14.60774 | 1.80712e-05 |
| 206 | 1237662500013670634 | 253.97184 | 25.16256 | 1.56238e-05 |
| 207 | 1237662198814671043 | 214.29213 | 10.57106 | 7.98167e-06 |
| 208 | 1237660960792772640 | 140.34447 | 34.45554 | 4.57369e-06 |
| 209 | 1237654874824048784 | 188.03774 | 64.10488 | 7.47752e-06 |
| 210 | 1237664336634773574 | 156.16240 | 32.65957 | 6.94503e-06 |
| 211 | 1237665372259680469 | 239.66761 | 19.47296 | 2.93035e-06 |
| 212 | 1237658423024222319 | 176.73398 | 6.87771 | 2.86194e-06 |

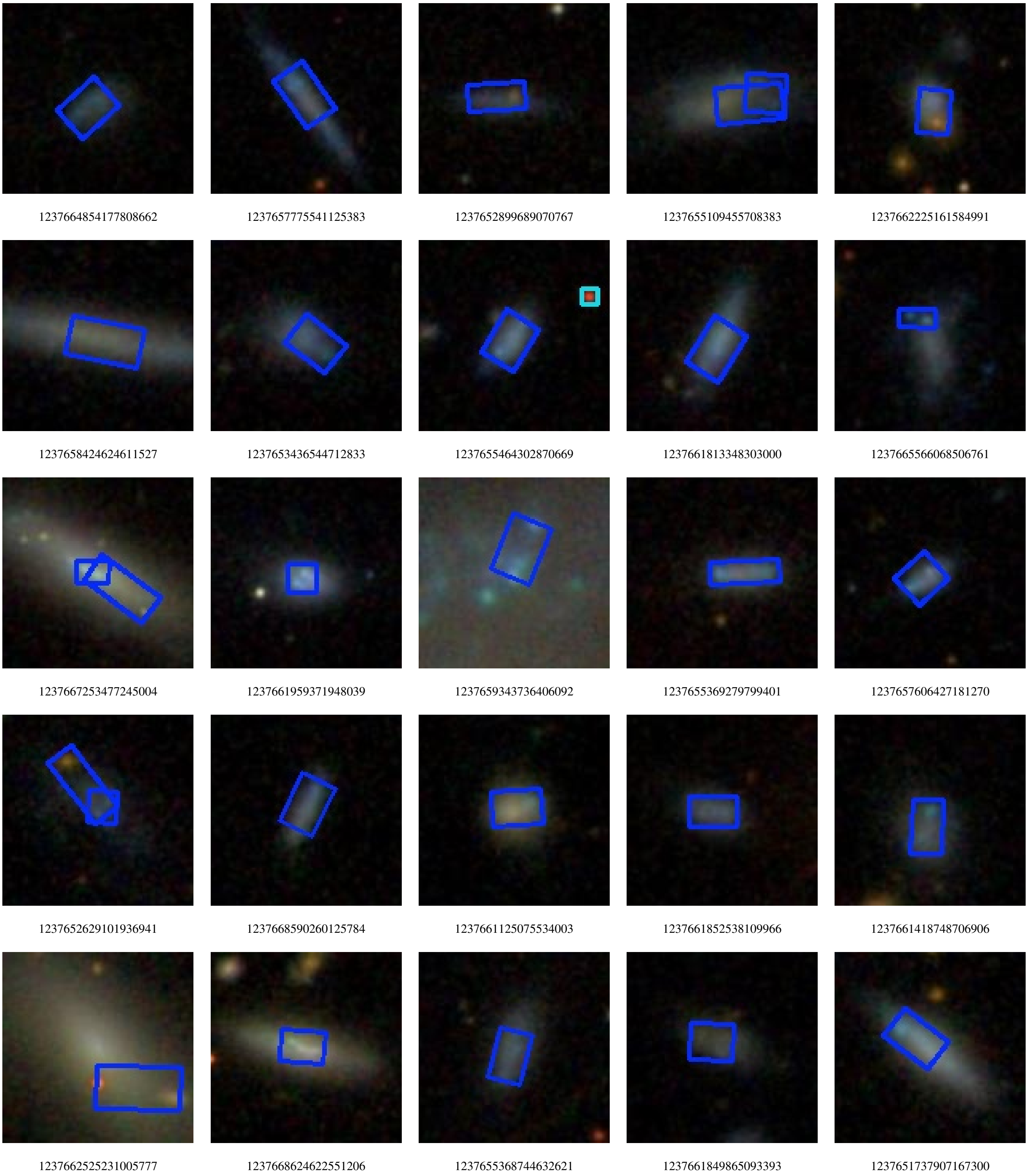

**Figure 1.** SDSS colour-composite image cutouts of the 212 sub-kiloparsec dual-nucleus candidates ($s_{\rm GOTHIC} \leq 1$ kpc) listed in Table 1, shown in the same order. Each panel is a $120 \times 120$-pixel cutout ($0.3''$ pixel$^{-1}$; $36'' \times 36''$ on the sky) labelled by its SDSS `objID`. At these separations the two nuclei are unresolved in SDSS imaging (Section 4.3).

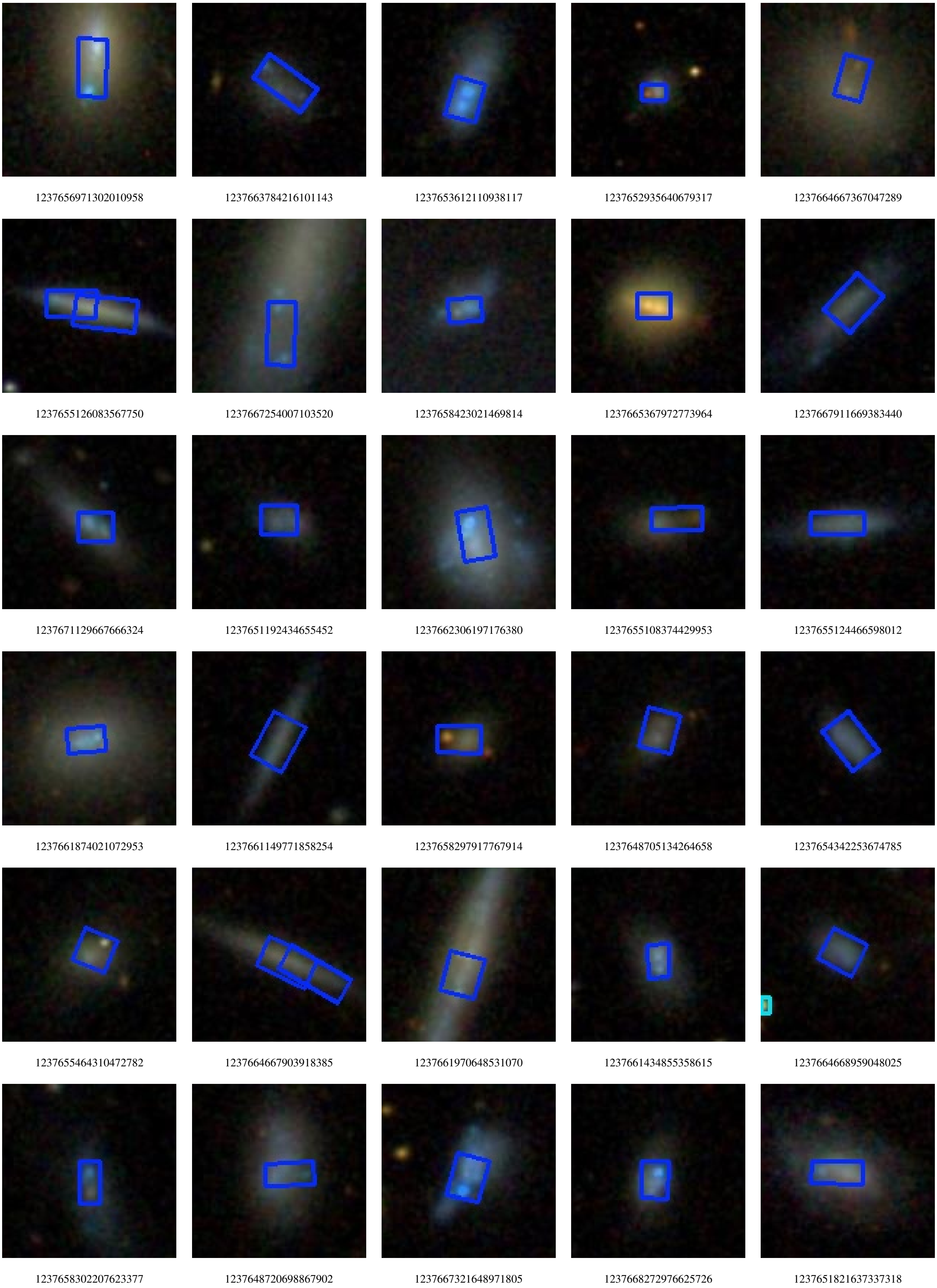

**Figure 2.** SDSS image cutouts of the sub-kiloparsec dual-nucleus candidates (Fig. 1, continued).

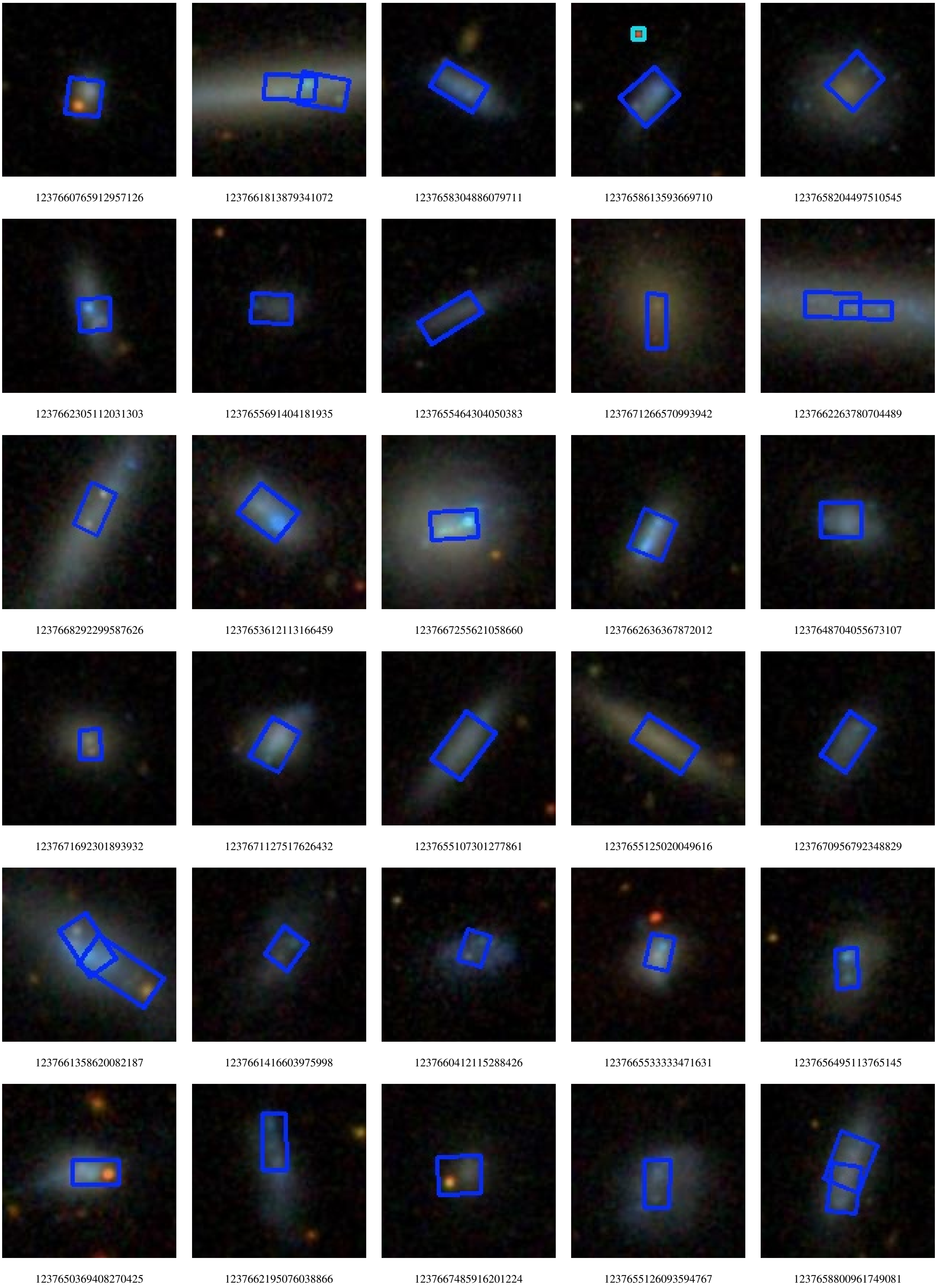

**Figure 3.** SDSS image cutouts of the sub-kiloparsec dual-nucleus candidates (Fig. 1, continued).

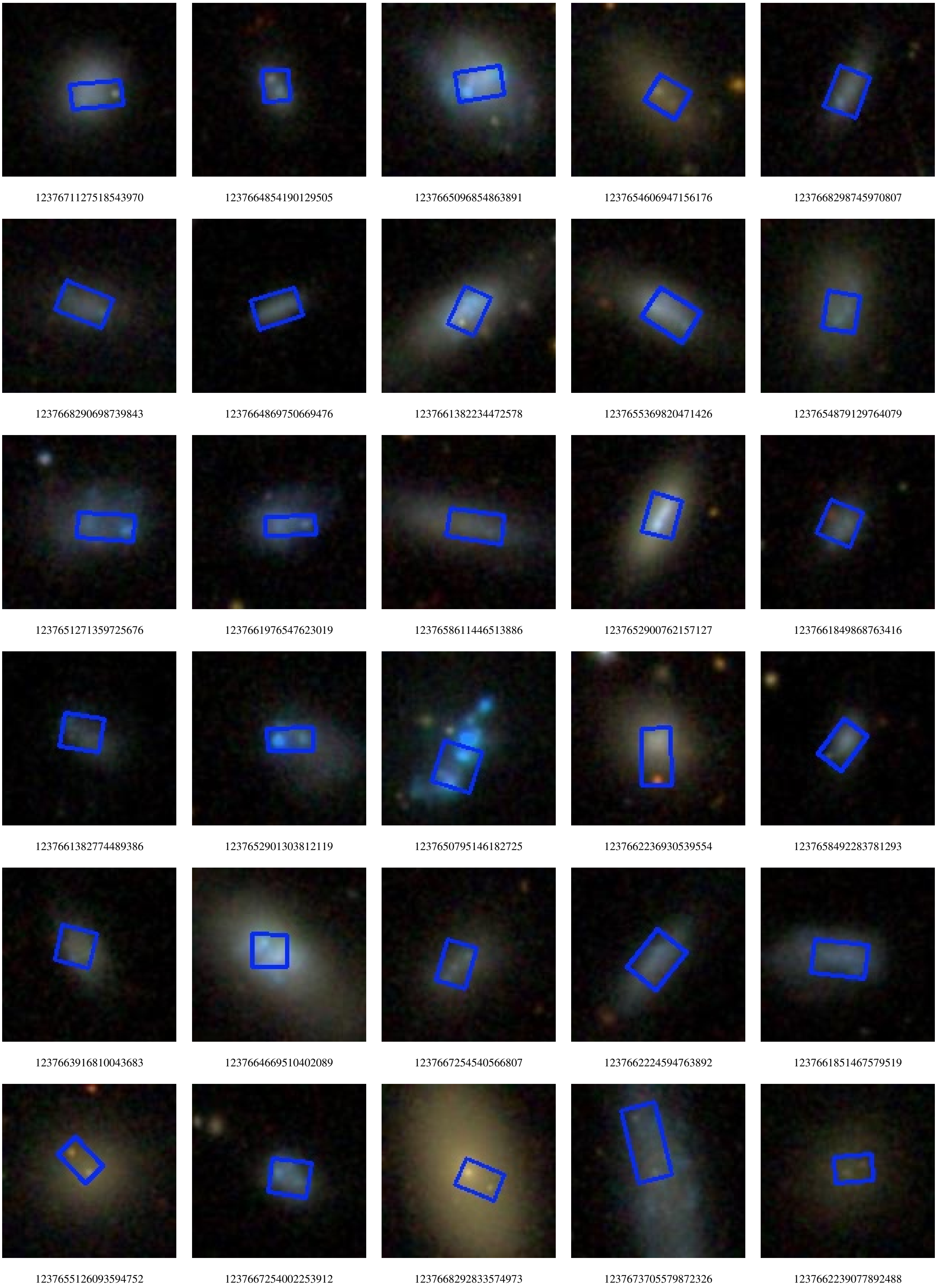

**Figure 4.** SDSS image cutouts of the sub-kiloparsec dual-nucleus candidates (Fig. 1, continued).

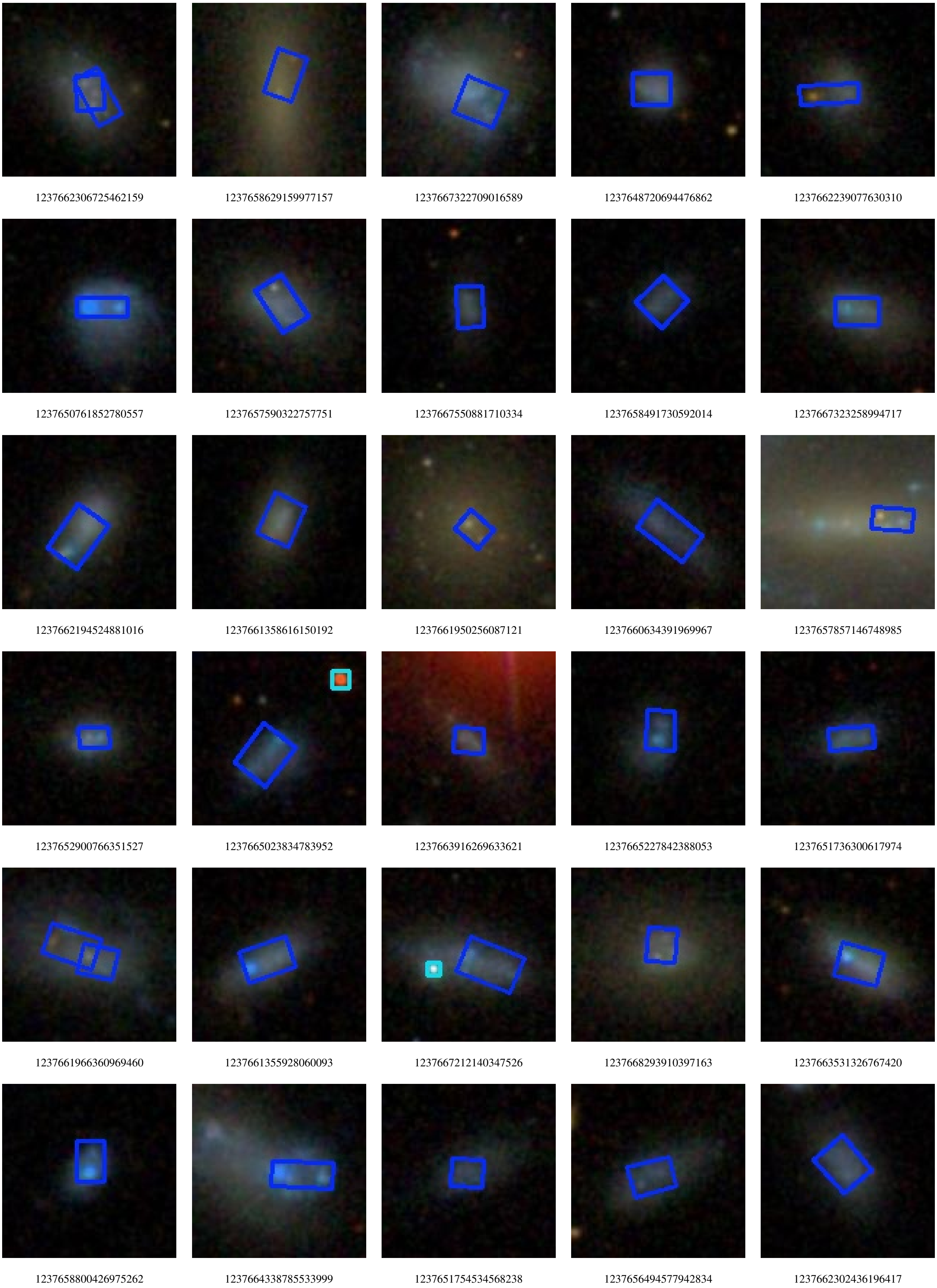

**Figure 5.** SDSS image cutouts of the sub-kiloparsec dual-nucleus candidates (Fig. 1, continued).

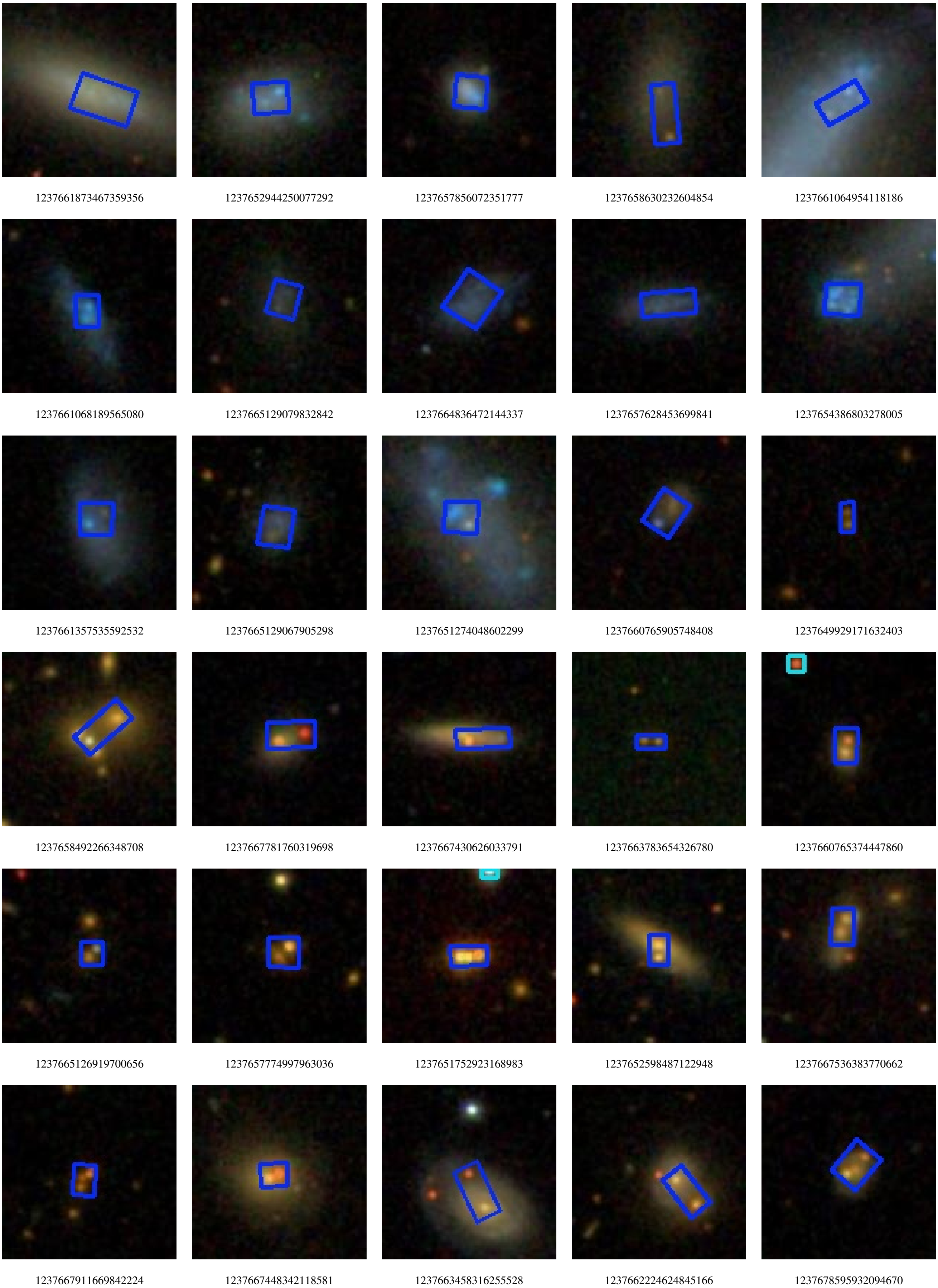

**Figure 6.** SDSS image cutouts of the sub-kiloparsec dual-nucleus candidates (Fig. 1, continued).

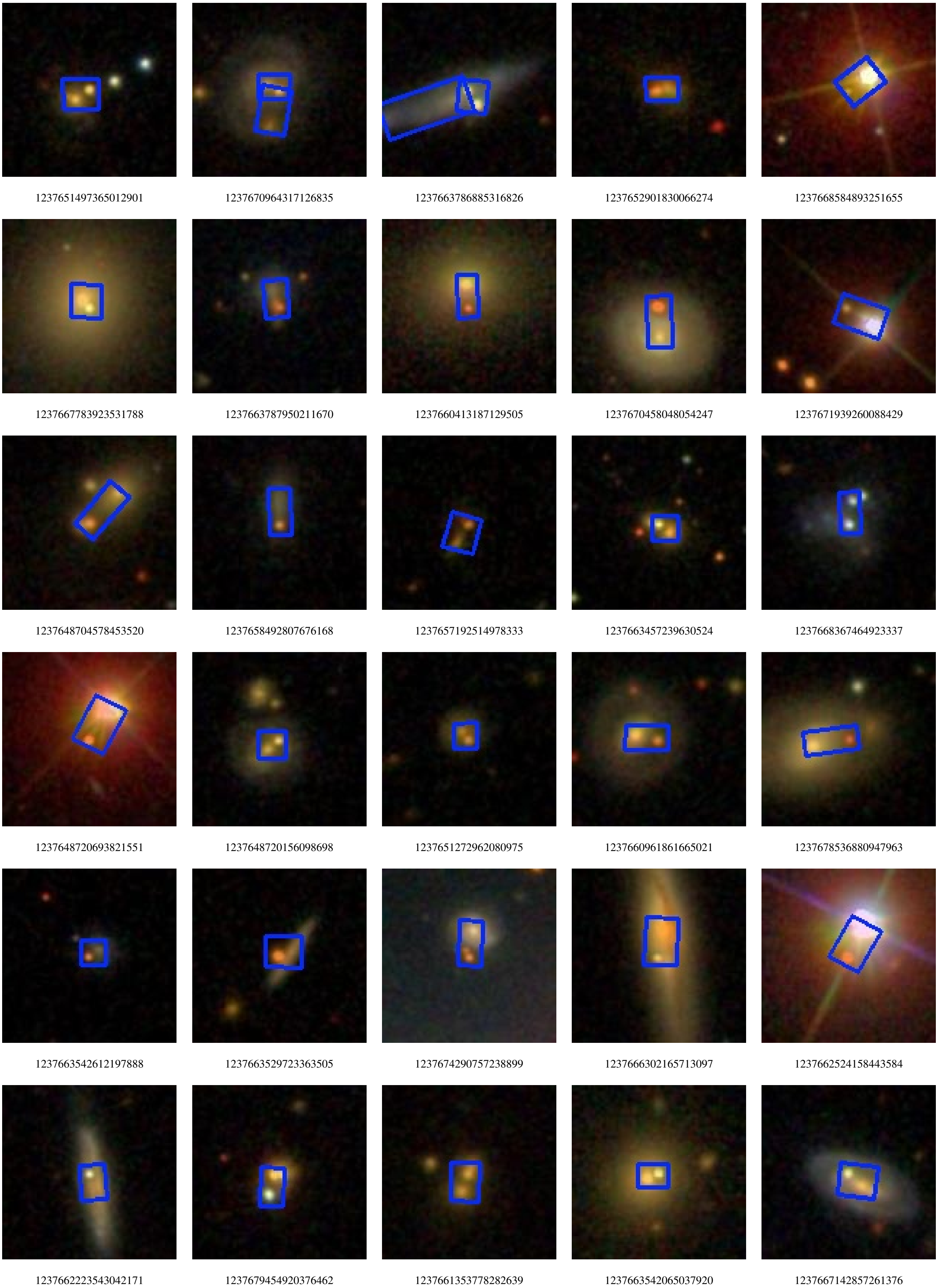

**Figure 7.** SDSS image cutouts of the sub-kiloparsec dual-nucleus candidates (Fig. 1, continued).

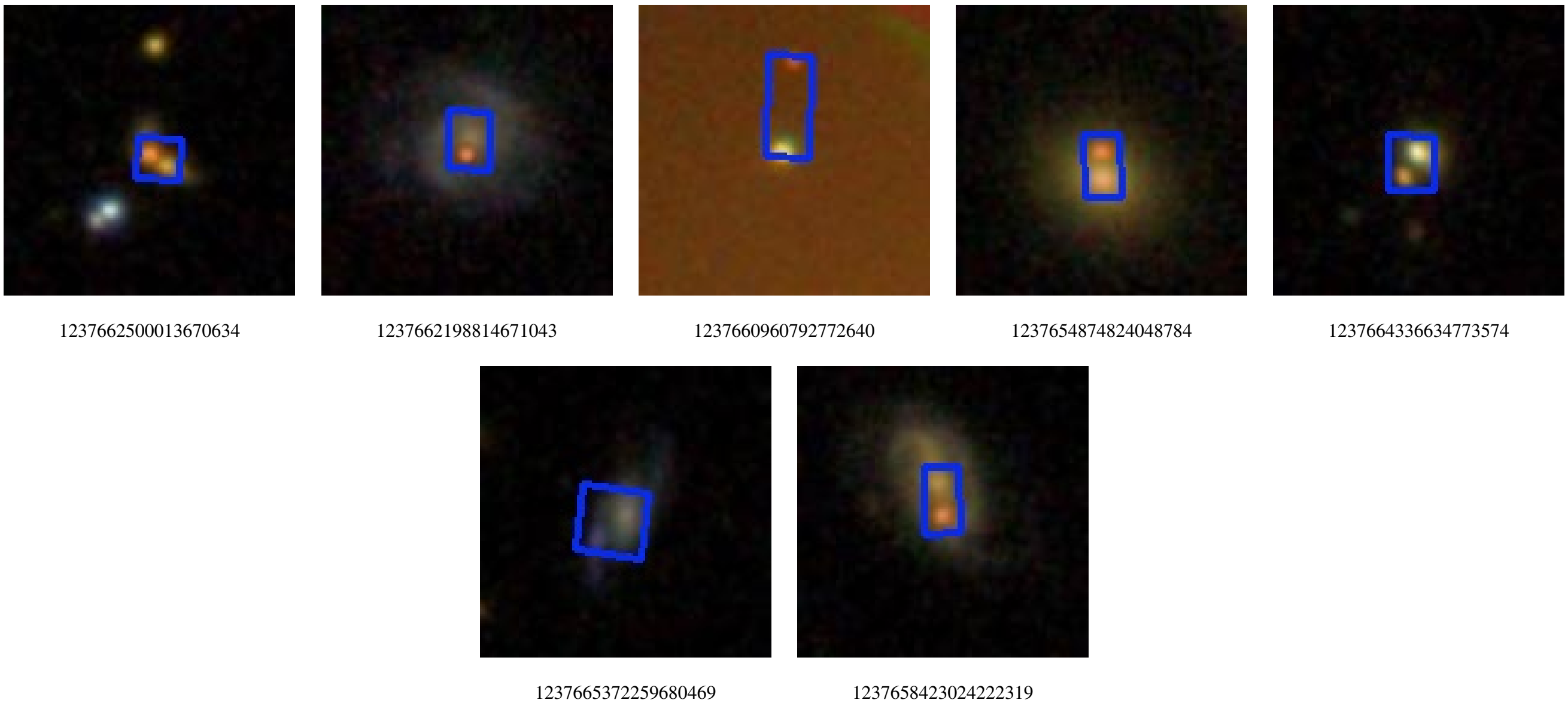

**Figure 8.** SDSS image cutouts of the sub-kiloparsec dual-nucleus candidates (Fig. 1, continued).